# A Novel Flow-induced Motion Energy Harvesting with Coupled Mechanism of Time-varying Stiffness and Passive Turbulence Control

Yongxi Wu[2], Hao Wu[1]*

[1] School of Mechatronic Engineering and Automation, Shanghai University, Shanghai, 200444, P.R. China.

[2] School of Mechanics and Engineering Science, Shanghai University, Shanghai, 200444, P.R. China.

**Abstract**

The ocean contains a substantial amount of energy, and the efficient harvesting of this energy holds significant importance. Drawing inspiration from the biomimicry of octopus tentacles, this study introduces a synergistic mechanism designed to optimize energy harvesting through flow-induced motions, integrating boundary layer modulation via passive turbulence control (PTC) with dynamic system stiffness adjustments via time-varying stiffness (SIN & Trapezoidal patterns). The implementation of PTC facilitates global stability by managing the local instabilities caused by variable stiffness, culminating in a highly effective energy harvesting capability. Our investigations summarize the requisite conditions for peak energy harvesting efficacy, notably within the SIN/80° PTC and Trapezoid/60° PTC arrangements, which have been demonstrated to double the efficiency of energy harvesting with up to 57%, alongside a reduction in initial harvesting frequency and an enhancement in both instantaneous power output and vibration amplitude. Furthermore, an energy transfer characteristic map has been compiled to illustrate the mechanism coupled between boundary layer modulation and time-varying stiffness. This research not only introduces novel perspectives but also stands as a significant stride in the realm of wide band and efficient energy harvesting in the ocean.

**Keywords**: Flow-induced Motion; Efficient Energy Harvesting; Coupled Mechanism; Boundary Layer Modulation; Time-varying Stiffness; Bionics

| Nomenclature | | | |
|---|---|---|---|
| $I$ | Moment of inertia | $D$ | Diameter of cylinder |
| $L$ | Length of cylinder | $U$ | flow velocity |
| $X$ | Displacement of cylinder | $U^*(2\pi U/(\omega_n D))$ | Reduced flow velocity |
| $f$ | Frequency | $A$ | Amplitude of cylinder |
| $K$ | Physical linear spring of stiffness | $\zeta$ | Physical constant damping ratio |
| $\omega_n$ | Natural frequency of constant stiffness in still water | $P$ | Power |
| $\omega_v$ | The frequency of time-varying stiffness | $A^*(A/D)$ | Dimensionless displacement amplitude |
| $X^*(X/D)$ | Dimensionless displacement | $f^*$ | Dimensionless frequency |
| $A_v/A^*$ | Dimensionless amplitude ratio of time-varying over constant stiffness | $f_v/f^*$ | Dimensionless frequency ratio of time-varying over constant stiffness |
| $P_v/P^*$ | Dimensionless power ratio of time-varying over constant stiffness | $\eta$ | Efficiency of energy harvesting |

## 1. Introduction

Many scholars have studied energy harvesting in-depth, but for the actual marine environment, there are some disadvantages of narrow energy harvesting frequency band and low efficiency. To solve such problems, the idea of boundary layer control is introduced and passive turbulence control (PTC) is one of it. PTC can achieve precise regulation of boundary layer separation. Thus, the flow field distribution and the force interaction between the fluid and solid are controlled, producing the effect on the vibrator dynamics finally. Many scholars have studied the influence of PTC material, height, location and other factors on flow field, vortex-induced vibration amplitude, frequency, energy harvesting efficiency, etc., which will be introduced in detail next.

Vortex-induced vibration is a commonly observed and destructive hydrodynamically excitation of elongated flexible structures exposed to fluid flows, such as ocean or river flows. Achenbach et al. [1], Guven et al. [2], Nakamura and Tomonari [3] had done many researches on the roughness which is uniformly distributed over the entire cylinder surface. Most of these studies aimed to determine the pressure distribution around the cylinder, flow separation, and Strouhal number characteristics. Few of these studies focused on the hydrodynamic excitation of the body. PTC trips flow separation and energizes boundary layer. Chang et al. [4] investigated the effects of roughness position, surface cover, and size on vortex-induced vibration (VIV), enhanced VIV, or galloping experimentally, proving that 16 degrees roughness coverage is effective in the range (10 degrees–80 degrees) by employing a passive turbulence control in the form of selectively distributed roughness strips on the surface of the circular cylinder. Allen et al. [5] testing the effect of four levels of surface roughness on VIV and resistance upstream of VIV and downstream cylinder found that surface roughness plays a very important role, not only in the drag coefficient, but also in the VIV response. If the surface roughness of the cylinder is smooth enough, the VIV almost completely disappears and the resistance coefficient is very low. Blevins and Coughran [6] also reported the response amplitudes of rough cylinders where the roughness was applied to the entire cylinder surface. They reported a 2-fold decrease in the amplitude of rough cylinders with a roughness / diameter value of 0.005.

In addition to the research on the structural properties of PTC itself, flow velocity, an external energy source variable, cannot be ignored. Vinod et al. [7] conducted the experimental work which identified the mechanism of the vortex-induced vibration (VIV) and the Reynolds number. The experiment belonged to the TrSL2 ($1.5\times10^3 <$ Re $< 3\times10^4$) (transition of the shear layer 2) system. They use smooth rectangular bar pairs of different thickness (1.6% -31% of cylinder diameter) attached to a circular cylinder, discovering that thicker strips led to higher VIV and galloping amplitudes, accompanied by an increase in steadiness within the transition regime, leading to earlier initiation of galloping, indicating capability for increased energy transfer even at the lower flow speeds. Previous work by Vinod and Banerjee [8] tested the effectiveness of PTC mechanisms in FIM augmentation in the TrSL2 Reynolds number regime, discovering that the cylinder attached with smooth strips reached higher amplitude, higher frequency galloping oscillations than those with rough strips. Ding et al. [9] investigated Flow Induced Motions (FIM) of a single, rigid, circular cylinder on end-springs for Reynolds number 30,000 < Re < 110,000, applying PTC in the form of roughness strips to enhance FIM and increase the efficiency of the VIVACE (Vortex Induced Vibration for Aquatic Clean Energy) converter in harnessing marine hydrokinetic energy, reaching an energy conversion efficiency of 37% in simulations and 28% in experiment. They also simulated the flow-induced motion of multiple cylinders with a steady and uniform flow control, considering four configurations of cylinders in series with a comparison of kinematic response and vortex at different locations [10].

With a large number of numerical simulation and experimental results of PTC-related variables, some researchers try to summarize and propose systematic conclusions that can have guiding value. Park et al. [11] proposed the map of "PTC-to-FIM" developed in previous work, revealing robust zones of weak suppression, strong suppression, hard galloping, and soft galloping, guiding the development of FIM suppression devices. PTC was revealed the potential to suppress or enhance FIM to various levels. Park et al. [12,13] applied passive turbulence control (PTC) in the form of selectively distributed surface roughness on a rigid circular cylinder on two end-springs, studying the FIM in the soft galloping and the two hard galloping zones identified in the PTC-to-FIM Map. The galloping range follows the VIV range thus expanding dramatically the FIM range, revealing the fundamentally different driving mechanism of VIV and galloping. They also experimentally studied the influence of the width of PTC covering, a single or multiple zones, and straight or staggered. Li et al. [14] investigated the effect of damping on FIO and power extraction for a converter with large turbulence stimulation (PTC), revealing the mechanism behind the variation of harnessing efficiency with inflow velocity and damping ratio, discussing the optimality criterion for the converter design. He et al. [15] experimentally discussed the influence of roughness of strips and thickness and decouples these two effects for augmented energy extraction, providing more operational stability, using these devices for energy harvesting perspective at low velocities (<1.0 m/ s).

Although a large number of previous studies have been done related to vortex-excited vibrational energy harvesting and the effect of PTC on it, there are few studies on how to realize more efficient and broadband energy harvesting under TrSL2 (transition of the shear layer) system considering both stiffness of the system and the effect of PTC. The aim of current work using cylinders is to identify the mechanism of amplified vortex-induced vibration within the Reynolds number range $1.2*10^4 <Re <2.4*10^4$(TrSL2).

In order to improve the energy harvesting ability of cylindrical structures in this dynamic environment, a bionic oscillator model with Variable Stiffness coupled with PTC System (VSPTCS) using numerical simulation method is introduced. And the cylindrical vortex-excited vibration energy harvesting characteristics with variable stiffness under zone of transition of the shear layer 2 are explored as shown in Fig.1.

First, the PTC and variable stiffness coupled energy harvesting model is introduced with the variable stiffness design bionic mechanism and relevant theoretical models described followed by the numerical simulation methodology and the validity of the numerical simulations. In the third part, the kinematic characteristics of the cylinders in terms of power, amplitude at different velocities, PTC positions, variable stiffness mechanisms, through the velocity cloud, vortex cloud, and surface pressure coefficient are analyzed. In the end, a discussion and summarization are made for the whole paper, revealing the energy harvesting process under effect of PTC and variable stiffness, proposing the energy transfer mechanism diagram.

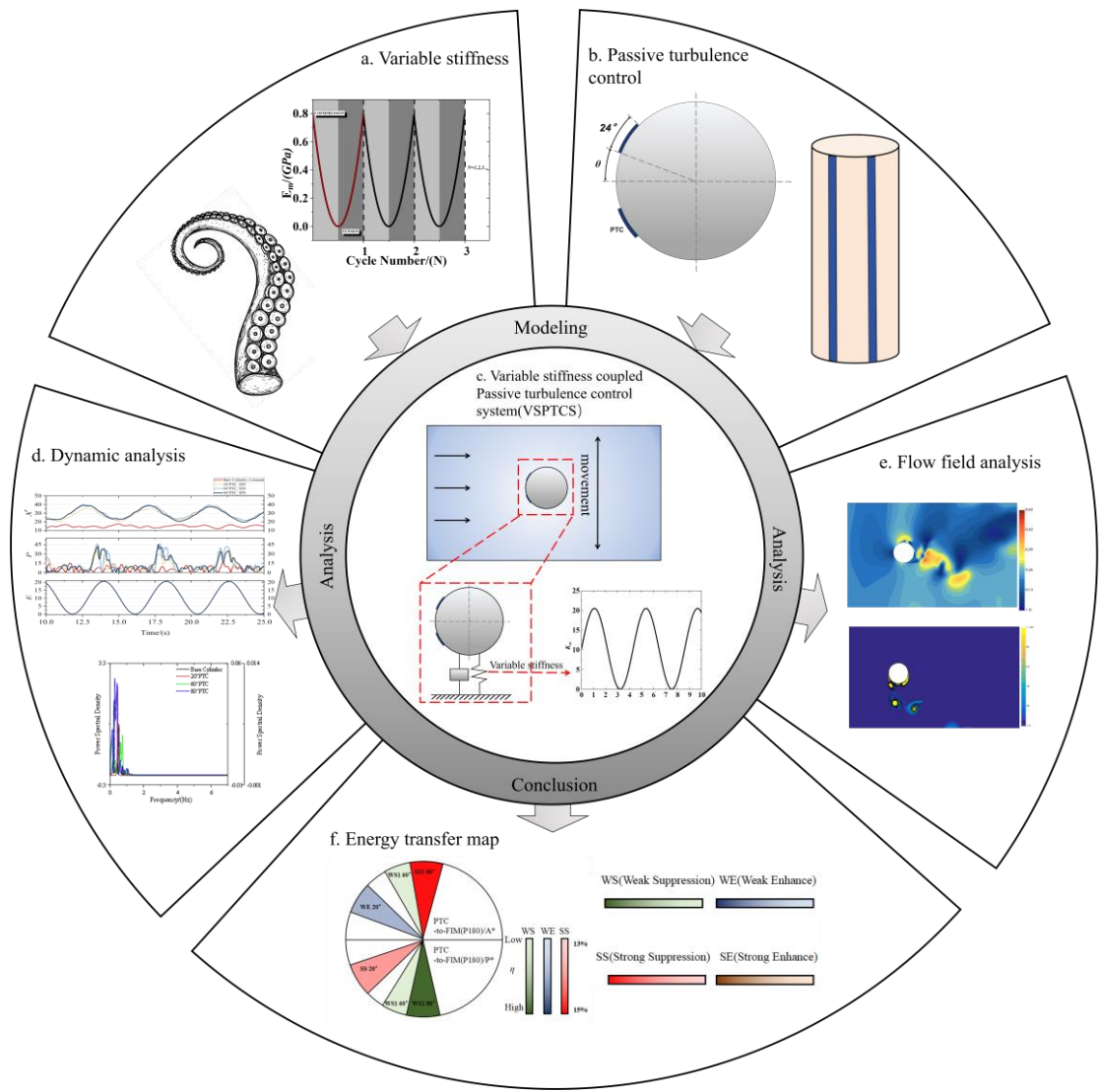

Fig. 1 Schematic diagram. (a) Biomechanical model with variable stiffness. (b) Passive turbulence control device. (c) Energy harvesting mechanism of variable stiffness coupled passive turbulence control. (d)Dynamic analysis. (e) Flow field analysis. (f) Analysis of energy transfer mechanism

## 2. PTC and Variable Stiffness Coupled Energy Harvesting Model

### 2.1 Single-Degree-of-Freedom Fluidic Vibration Model

This study continues the previous works of Wu et al. [16-19]. The mathematical model and experimental equipment used are exactly the same as before. Cylinders subjected to low-turbulence free surface flow undergo a broad spectrum of Reynolds numbers concerning VIV. Interaction with the airflow induces the creation and shedding of vortices around the body. These vortices, in turn, generate harmonic forces or momentum, initiating body movements referred to as FIM. To integrate the cylinder's dynamics into a dynamic simulation model, a mass damper-spring oscillator was structured using a linear second-order ordinary differential equation (ODE). This ODE elucidates the forces imposed on the cylinder by low turbulence.

$$(I+I_A)x'' + cx' + K_{vt}x = M(t) = M_{flow} \tag{1}$$

The displacement of the cylinder in the cross direction, represented as $X$, is influenced by various factors such as the righting moment stiffness ($K_{vt}$), torsional damping ($c$), and the force exerted by the flow on the cylinder ($M_{flow}$). The response characteristics of the cylinder interacting with the flow can be described through two key parameters: the response amplitude ($A$) and the response frequency ($\omega$). The dimensionless displacement is defined as $A^* = A/\alpha$, where $\alpha = \arctan(0.5D/L)$, with $D$ being the cylinder's diameter and $L$ being its length.

### 2.2 Numerical Simulation Model

#### 2.2.1 Numerical Computational Methods for Control Equations and Turbulence Model

Numerical simulation is performed in the range $1.2*10^4 < Re < 2.4*10^4$ and is carried out using the SST $k$-$\omega$ turbulence model by solving the non-constant Reynolds-averaged Navier-Stokes (URANS) equations. The SST $k$-$\omega$ turbulence model is used for the calculations and the shear stress transport (SST) formulation combines the advantages of $k$-$\omega$ and $k$-$\varepsilon$. The use of $k$-$\omega$ inside the boundary layer and $k$-$\varepsilon$ in the free stream avoids the problem that the model is too sensitive to the turbulence properties of the inlet free stream, i.e., initial value sensitivity. This also allows the SST $k$-$\omega$ model to simulate accurately in unfavorable pressure gradients and separated flows in a way that is appropriate for the main problem studied. While it is true that the SST $k$-$\omega$ model produces excessive errors in regions of high normal strain. However, this tendency is much less pronounced than for the normal $k$-$\varepsilon$ model. In particular, for boundary layers subject to reverse pressure gradients, the model is validated for fluid-structure interaction problems, free shear layers, zero or reverse pressure gradients in boundary layers.

The Reynolds-averaged Navier-Stokes equations and the continuity equations are as follows

$$\rho \frac{\partial u_i}{\partial t} + \rho \frac{\partial (u_i u_j)}{\partial x_j} = \frac{\partial p}{\partial x_j} + \frac{\partial}{\partial x_j}[\mu(\frac{\partial u_i}{\partial x_j} + \frac{\partial u_j}{\partial x_i} - \frac{2}{3}\delta_{ij}\frac{\partial u_i}{x_i})] + \frac{\partial}{\partial x_j}(-\rho \overline{x'_i x'_j})$$
(2)

$$\frac{\partial \rho}{\partial t} + \frac{\partial}{\partial x_i}(\rho u_i) = 0$$
(3)

In addition, the SST $k$-$\omega$ model adds a cross-diffusion term to the $\omega$ equation and the mixing function to provide a model that behaves more favorably in both the near-wall and far-field regions. The transport equation for SST $k$-$\omega$ is given below:

$$\frac{\partial}{\partial t}(\rho k) + \frac{\partial}{\partial x_i}(\rho k u_i) = \frac{\partial}{\partial x_i}[\Gamma_k \frac{\partial k}{\partial x_j}] + G_k - Y_k + S_k$$
(4)

$$\frac{\partial}{\partial t}(\rho \omega) + \frac{\partial}{\partial x_i}(\rho \omega u_i) = \frac{\partial}{\partial x_i}[\Gamma_\omega \frac{\partial \omega}{\partial x_j}] + G_\omega + D_\omega - Y_\omega + S_\omega$$
(5)

The turbulent kinetic energy production term $G_k$ is computed from the $k$-$\varepsilon$ turbulence model. the terms $\Gamma_k$ and $\Gamma_\omega$ are the main difference between the $k$-$\omega$ turbulence model and the $k$-$\varepsilon$ turbulence model, i.e., the effective diffusion of $k$ and $\omega$. $Y_i$ and $Y_\omega$ describe the dissipation of $k$ and $\omega$, respectively. the derivation of the $S$ term is also described as a user-defined source term.

**2.2.2 Bionic and Flow Field and Oscillator Model**

Octopus, as a mollusk, has unique biomechanical properties that allow it to have variable stiffness characteristics, and exhibits excellent performance in mechanical adaptation and motion control. The physiological characteristics of the variable stiffness of octopus foot to design a bionic mechanism is drawn with the aim of realizing broadband and efficient energy harvesting in random marine environments.

Mazzolai et al. [20] investigated the biomechanical characteristics of the elastic and tensile-compressive stress/stretch curves of octopus tentacles. The stress-stretch curves of octopus tentacles are very similar to those of hyperelastic materials, where the stress increases in a nonlinear fashion with increasing stretch which is shown in Fig.2. As shown in Fig.3, a statistical approach to the data is used and fitted the results to infer analytical equations for the change in stiffness versus deformation. These fits are not from the theoretical equations of hyperelasticity, but represented analytical fits to the experimental data obtained. Using the fitted function that relates the stress to the deformation, the stiffness parameters can be deduced from the fitted point-by-point angular coefficients. In the ideal case without considering factors such as fatigue, and assuming that the linkage is subjected to N tensile-compression experiments, the obtained variation rule of $E$ with N/time is shown in Fig.4.

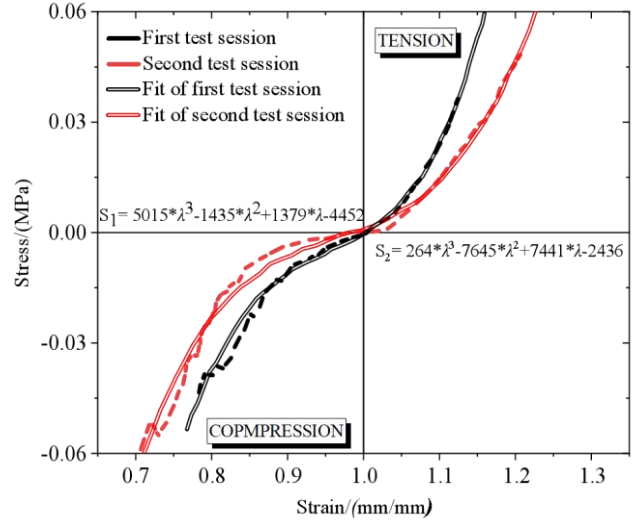

Fig. 2 Tensile-compressive stress/stretch curves of octopus tentacles

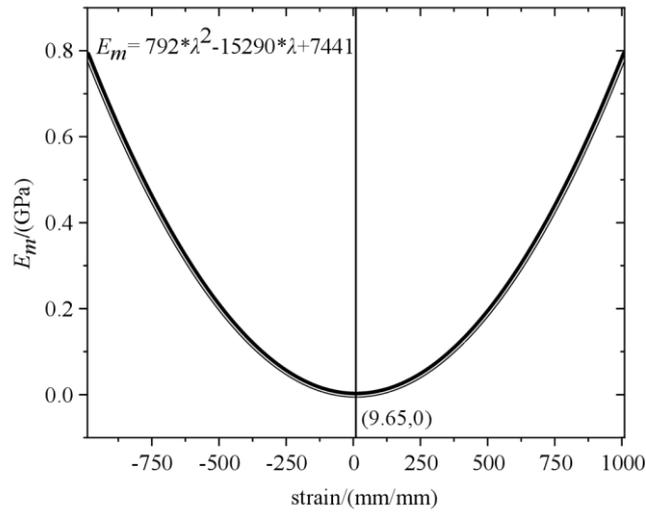

Fig. 3 Analytical equations for stiffness change and deformation

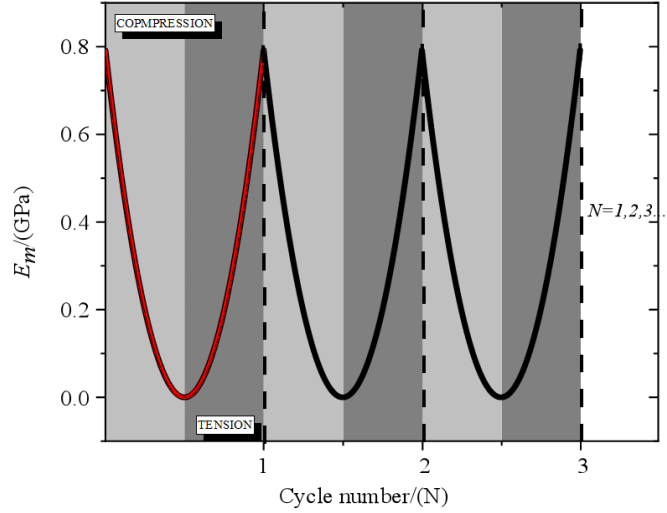

Fig. 4 Stiffness time histories under cyclic loading

The time-varying stiffness model herein derives from the dynamic variability observed in the stiffness pattern of octopus legs, serving as a replacement for the sinusoidal (SIN) oscillations typically generated during cyclic stretching and compression. SIN and trapezoid models are introduced to emulate this variability. Furthermore, the suction cups located on octopus tentacles significantly influence the boundary layer characteristics of the surrounding flow field during propulsion. Leveraging this bionic mechanism, the concept of boundary layer flow control is introduced through the application of sandpaper with a wide of 0.5 inch as a symmetrically positioned boundary layer flow control device along both sides of the stationary point on the cylindrical surface of the tentacles.

In the case of the variable stiffness setup, the amplitude of the time-varying stiffness (denoted as $K_0$) remained constant in the experiments conducted for this study. $K_0$ represents the dimensionless stiffness under the experimental constant stiffness condition (refer to Table 2). In addition, although some energy needs to be input in order to achieve variable stiffness, this part of the energy is very small and can be ignored compared to the energy generated by the device

Table 2 Two forms of variable stiffness

| Item | Pattern | Parameter | Value |
|---|---|---|---|
| SIN |  | $K_{vt}=K_0/2\times\sin(m\times\omega_n\times Time)+10.27$ $\omega_n$ the natural frequency of constant stiffness | $K_0$=20.53 $m$=0.50/0.75/1.00/1.50 $\omega_n$=2.92 |

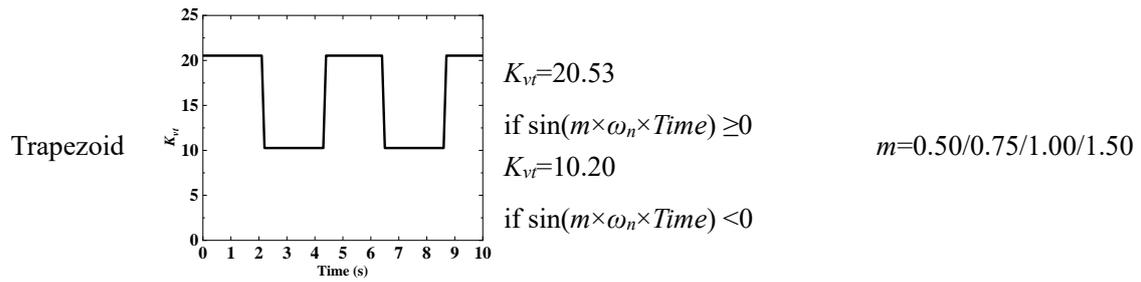

| Trapezoid | $K_{vt}=20.53$ if $\sin(m \times \omega_n \times Time) \geq 0$ $K_{vt}=10.20$ if $\sin(m \times \omega_n \times Time) <0$ | $m=0.50/0.75/1.00/1.50$ |

The introduced boundary layer flow control device holds the capability to transition the laminar shear layer into turbulence, effectively managing boundary layer separation. This transformative action aims to optimize motion dynamics and enhance energy harvest efficiency within the unpredictable oceanic environment. The degree of roughness of the boundary layer flow control device is characterized by the designated strip denoted as P180 (refer to Table 3), wherein a higher P value corresponds to a smoother surface texture. Fig 5 illustrates the development process of the bionic model. The tentacle of the octopus is modeled as a cylindrical structure with variable stiffness by incorporating a three-layer tissue configuration where $E(\lambda)$ represent elasticity modulus of tissue with differing elasticity modulus. The suction cups on the octopus tentacle are represented as PTC (Positive Turbulence Control) structures.

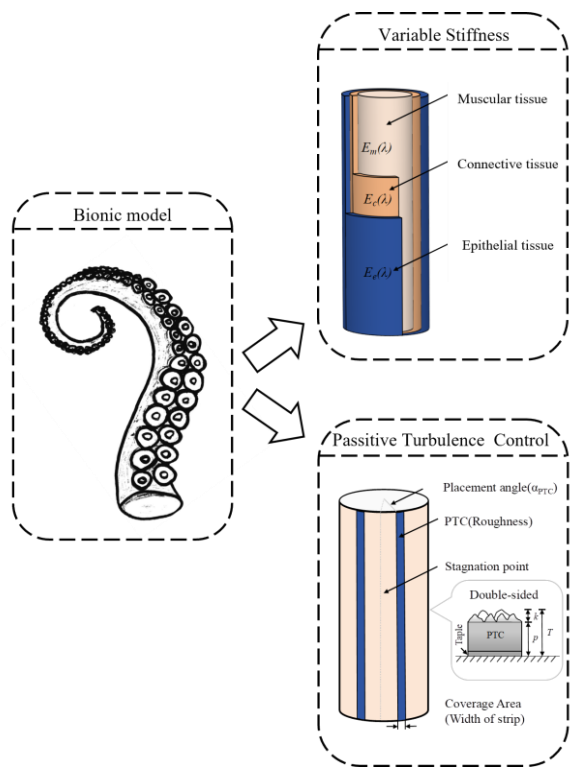

Fig. 5 Bionic passive turbulence control devices (PTC)

Table 3 PTC related structural parameters

| ISO/FEPA grit designation | Average particle diameter(μm) | Thickness of backup paper, $p$(μm) | k/D | (k+p)/D |
|---|---|---|---|---|
| P180 | 82 | 492 | $136\times10^{-5}$ | $952\times10^{-5}$ |

Investigating the impact of PTC position on VIV displacement, vibration frequency, and harvested energy, three sets of PTC positions are established with angles of 20 degrees, 60 degrees, and 80 degrees. The 20 degrees angle corresponds to a location where previous studies identified the optimal energy harvesting effect (Xu et al. [20]). The 60 degrees angle represents the position near boundary layer separation, while the 80 degrees angle corresponds to a location where the boundary layer has already separated. These three positions symbolize pre-separation, separation, and post-separation conditions, respectively as shown in Fig.6.

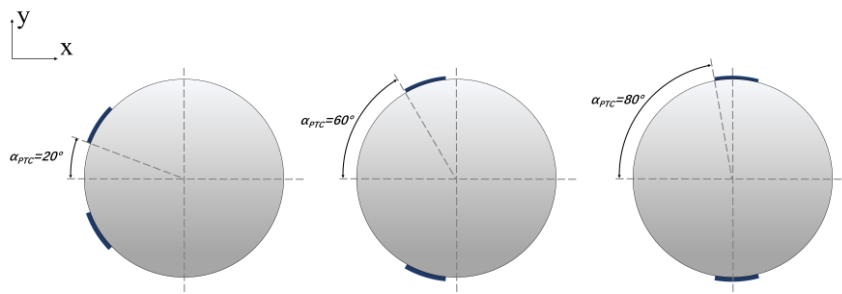

Fig.6 Different PTC distribution

The numerical simulation flow field is set up as shown in Fig.7, the inlet is the velocity inlet, the outlet is the pressure outlet, and the two symmetry surfaces are set up as the velocity inlet parallel to the velocity inlet in the same direction to simulate the real marine environment.

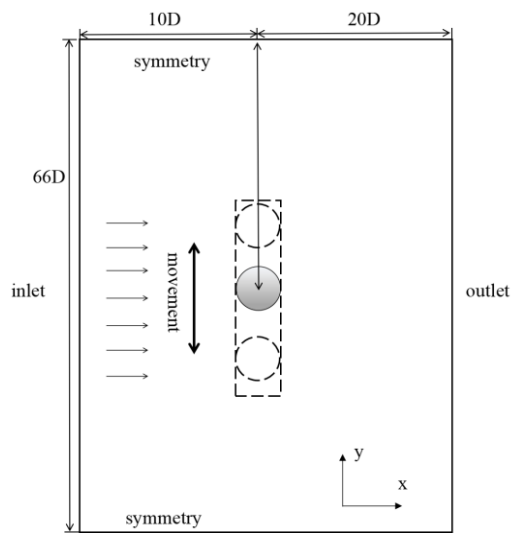

Fig.7 Computational domain

For the mesh setup as shown in Fig.8, in order to accurately simulate the force between the fluid and the cylinder during the single-degree-of-freedom cylindrical motion and the vortex shedding phenomenon after the cylinder, local encryption is carried out on the passing path of the cylinder and around the cylinder, and a reasonable boundary layer is set up in the vicinity of the PTC device, so that the value of y plus around the cylinder is about 1.0 in the final simulation results. The total number of mesh elements is 500,000, with the smallest mesh size outside the boundary layer being $3.15 \times 10^{-4}$ m.

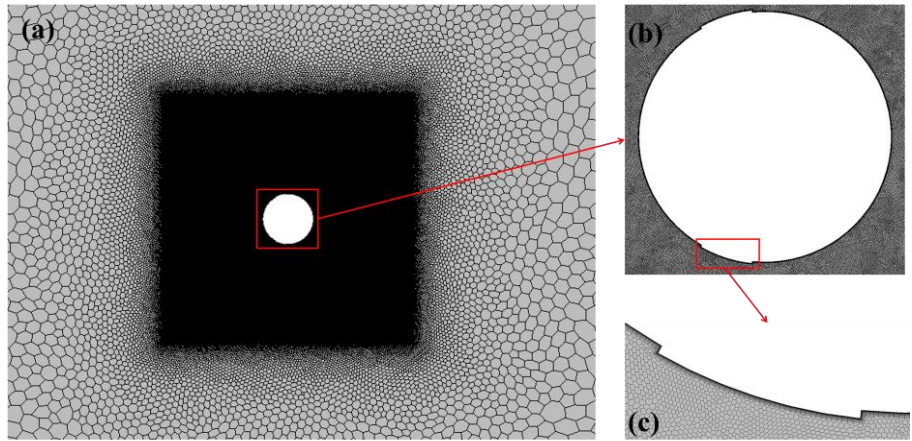

Fig.8 Mesh division. (a)Overall mesh. (b)Mesh around the cylinder. (c)Boundary layer

The oscillator setup comprises a rigid cylinder capable of rotational movement about a substrate-mounted hinge, coupled with a torsion spring. Key components of the oscillator setup include: (a) a rigid cylinder with dimensions specified by diameter D (6.032 cm) and length L (1.524 m); (b) a linear spring with a stiffness coefficient K = 31.281 Nm/rad; and (c) a linear viscous damping system characterized by a constant damping ratio $\zeta = 77.432 \times 10^{-3}$ as shown in Fig.9.

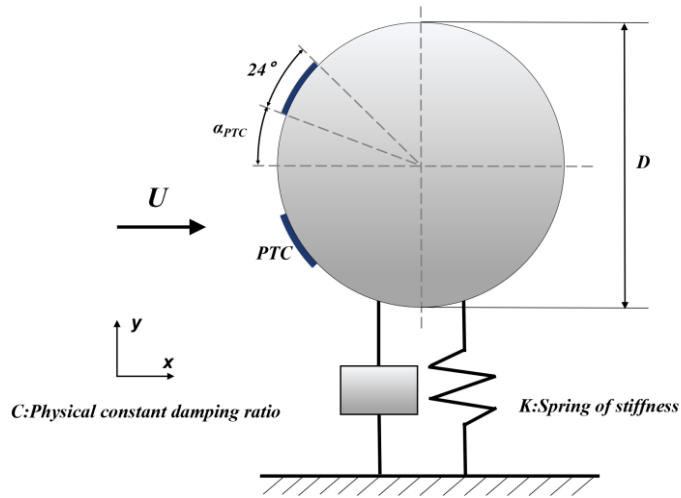

Fig. 9 Cylinder spring damping system with $c$ (physical constant damping ratio) and $K$ (spring of stiffness)

In order to verify the accuracy of the numerical simulation method, the obtained numerical simulation data are compared with the experiment. Physical testing of time-varying stiffness currently poses technological challenges. Consequently, experimental investigations were conducted utilizing a mass-damper-spring oscillator configured with a constant stiffness setup to gather essential motion and energy metrics. The experimental setup facilitated unrestricted movement of one side of the cylinder, measuring 1.0 m in width and 1.17 m in depth. Constructed from transparent Plexiglas, the setup allowed for visual inspection and analysis of both the experimental configuration and outcomes. Motion tracking of the cylinder was accomplished using an inclinometer affixed to the oscillator's bottom. To prevent any interference with the cylinder's motion, the cables connecting the inclinometer to the National Instruments data acquisition hardware were securely fastened to the top frame of the channel. In this experimental setup, data from the inclinometer were sampled at a rate of 30 Hz, without the application of any filters.

The natural frequency of the cylinder in still water is 2.923 rad/s as measured experimentally. By applying curve fitting, $\zeta=r/\omega_n=c/2(I+I_A)$, the damping ratio was calculated by analyzing the exponential decay coefficient, where parameters are shown in more detail in Table 4.

Table 4 Oscillator system particulars

| Item | Value |
| --- | --- |
| $I$ (Moment of inertia) [kg·m$^2$] | 5.867 |
| $D$ (Diameter of cylinder) [cm] | 6.032 |
| $L$ (Length of cylinder) [m] | 1.524 |
| $K$ (physical linear spring of stiffness) [N·m/rad] | 31.281 |
| $\zeta$ (physical constant damping ratio) | 77.432×10$^{-3}$ |
| $\omega_n$ (natural frequency in still water) [rad/s (Hz)] | 2.923(0.465) |
| $\mu$ [N·s/m$^2$] | 1.002×10$^{-3}$ |
| $\upsilon$ [m$^2$/s] | 1.000×10$^{-6}$ |
| $\rho$ [kg/m$^3$] | 998.207 |
| Temperature [°C] | 17.5-20.5 |

Comparing experimental vibration amplitudes between bare cylinders and those with varied PTC positions, alongside corresponding flow velocities, aims to validate the accuracy of numerical simulations as shown in Fig 10. The numerical simulations for the four cases exhibit high accuracy, with errors predominantly below the 5% threshold. This affirms the simulations' accuracy in harvesting the system's behavior under conditions of constant stiffness. In simulating time-varying stiffness conditions, the model remains constant, with considerations given to variations in stiffness. Hence, the variable stiffness simulation, based on the RANS model, is deemed reliable.

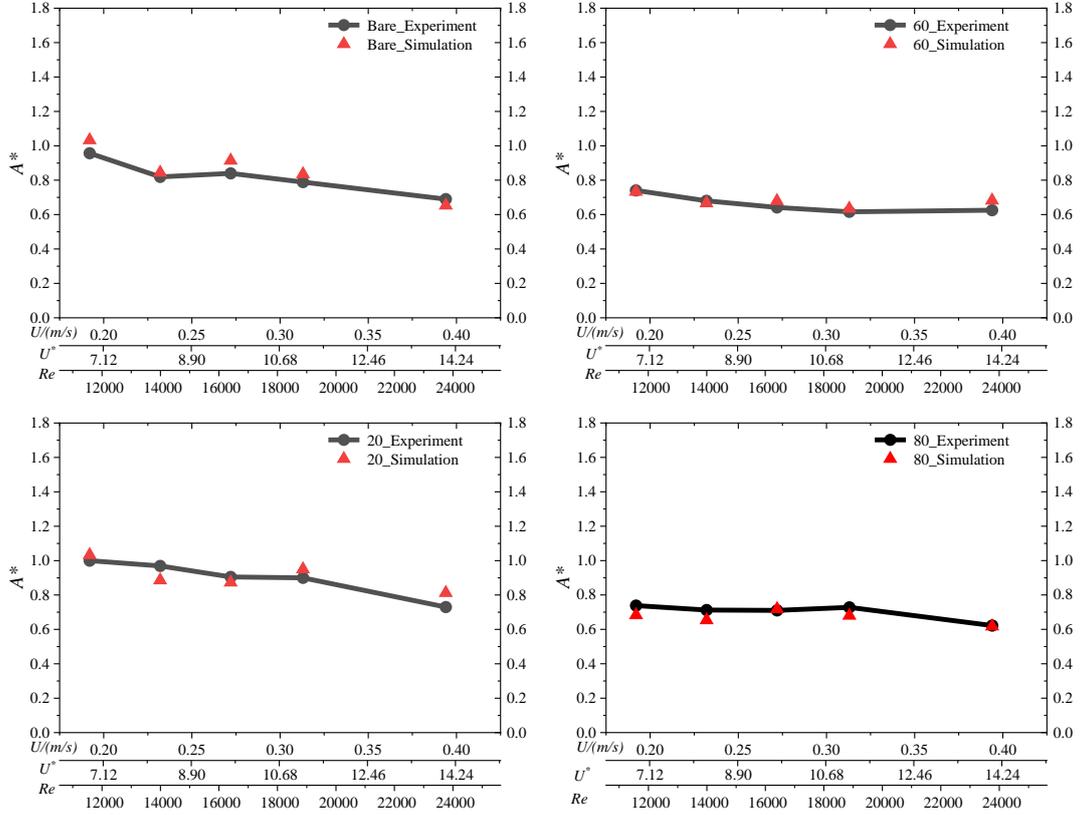

Fig. 10 Comparison between experimental values with the numerical simulation of amplitudes of bare cylinder and PTC at different positions

**2.3 Mathematical Modeling of Energy Transfer in Water**

As energy dissipates in fluid-structure interactions, only a fraction of the energy lost by the water is transmitted to the body. The energy transfer from the water to the oscillator can be dissected into two distinct components: the kinetic energy of the body and the dissipated energy. The damping coefficient, represented by $c$, emerges as a crucial parameter shaping the dynamics of the energy transfer process.

$$c = c_{kinetic} + c_{dissipated} \tag{6}$$

where $c_{dissipated}$ represents the damping associated with losses in the transmission system, and $c_{kinetic}$ represents the damping that converts the water energy into the mechanical energy of the body. Then, the mechanical power of the body can be expressed as

$$P = \frac{1}{2}cA^2\omega_{osc}^2 \tag{7}$$

where, $A$ is motion amplitude of the body, and $\omega_{osc}$ is angular frequency of oscillation, which are the statistics of the time history x(t). Uniform flow energy per transverse width is defined as

$$\boldsymbol{P}_{Fluid} = \frac{1}{2}\rho U^3 L(D+2A) \tag{8}$$

where, $\rho$ is water density, $U$ is flowing velocity, and $L$ is body length. The Betz limit, which corresponds to 59.26% (=16/27), represents the theoretical maximum power that can be extracted from an open flow. The power harvesting efficiency, denoted as $\eta$, can be calculated as follows:

$$\eta(\%) = \frac{P}{P_{Fluid}(BetzLimit)} \times 100 \tag{9}$$

**3. Results and Analysis**

**3.1 PTC Influence on Flow Field**

The following four figures show the dimensionless displacement and power time-varying curves of the bare cylinder and three different PTC positions and the time-varying screenshots of the flow field taken at four moments in order to analyze the variation of the flow field. The four moments are selected only to cover the different kinematic states to the maximum extent possible in one period and to compare in the same point of time, since this part of the study does not involve variable stiffness.

Bare cylinder as shown in Fig.11. For the bare cylinder, the unadorned cylinder's displacement exhibits a not particularly obvious periodicity over time, characterized by a notably high working frequency. Remarkably, during moment 'a,' the power output reaches its minimum concomitant with the peak displacement. The vortex shedding phenomenon presents a symmetrical disposition relative to the velocity field, featuring four vortices emanating from the distal terminus of the flow field. Notably, two vortices undergo detachment and subsequent coalescence, while another vortex disengages from the cylinder at this juncture.

Transitioning to moment **b**, the cylinder is traversing from a trough to a crest, accompanied by an elevated velocity. The vortex descends below the horizontal axis due to the cylinder's motion, leading to a superimposed wake around the cylinder. This upward velocity engenders a robust vortex shedding, propelling the vortices forcefully to the rear of the slope, where all four vortices become intricately implicated.

At moment **c**, the cylinder, having recently traversed the trough, ascends, instigating a drift phenomenon among previously shed vortices due to inertia. The wake undergoes dispersion, aggregating both formed and nascent vortices.

Advancing to moment **d**, a further extension of the preceding **c** moment, the vortex structure maintains relative symmetry. However, the intensity of vortex shedding remains pronounced. The wake aligns parallel to the incoming flow, and simultaneously, two vortices coalesce, facilitating downstream movement. This nuanced analysis provides a comprehensive understanding of the dynamic interplay between displacement, power, and vortex shedding in the context of the unadorned cylinder's behavior.

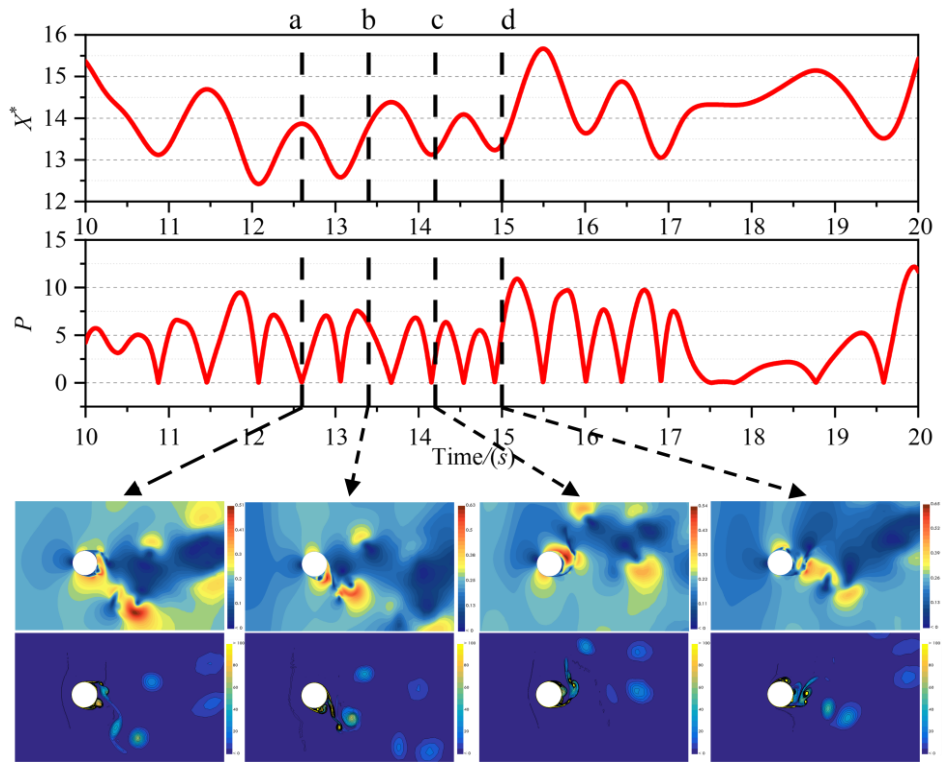

Fig. 11 Time-varying analysis of flow field for bare cylinder with constant stiffness

20 degrees PTC cylinder as shown in Fig.12. The alteration in the 20 degrees cylinder displacement exhibits a high degree of regularity, contrasting with the intricate and disorderly nature of fluid forces acting upon it, resulting in erratic power fluctuations. During moment **a**, the cylinder descends to its trough, corresponding to the minimum displacement and power.

Transitioning to moment **b**, as the cylinder ascends from the trough to the peak of the wave, the heightened velocity leads to an extreme power point. The vortex street, influenced by inertia, maintains a relatively symmetrical configuration during this peak, and subsequent shedding results in a downward drift of vortices around the cylinder.

At the moment **c**, the cylinder achieves its apex displacement, and the vortex shedding intensifies obliquely to the rear, exhibiting a pronounced and relatively symmetrical pattern reminiscent of vortex shedding around a stationary cylinder.

Subsequently, at the moment **d**, as the cylinder descends to the trough, the power diminishes, indicating a braking state, with a less intense vortex shedding. Despite the reduced shedding intensity, the flow field beneath the cylinder experiences accelerated speed in the direction of flow.

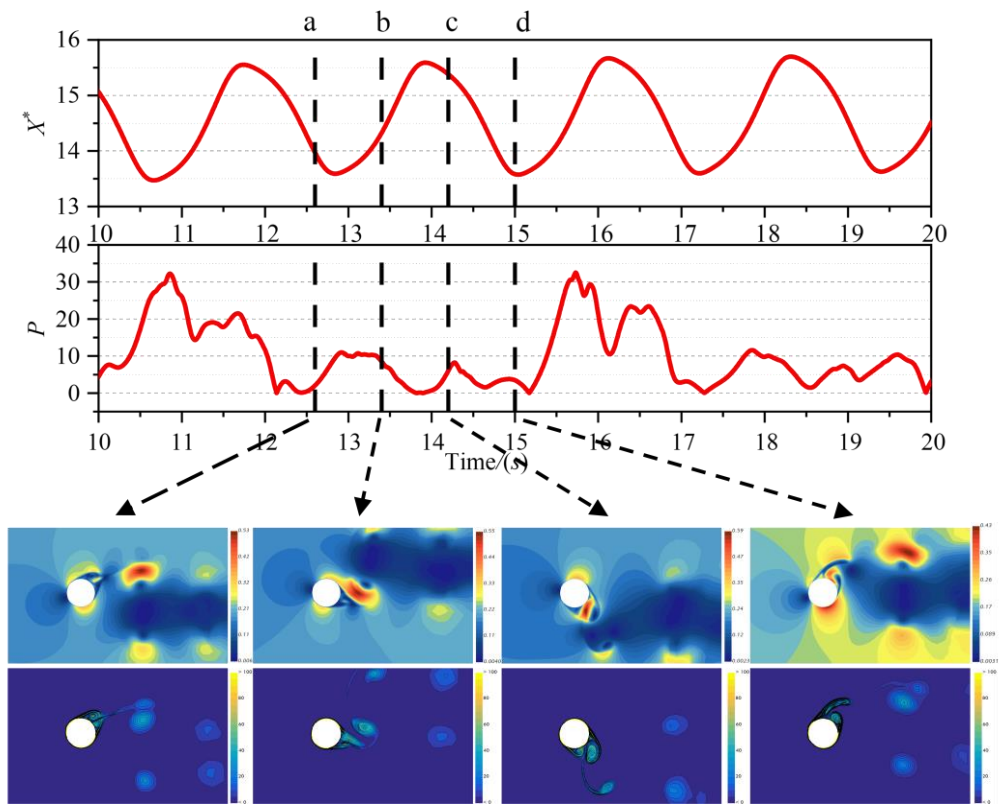

Fig. 12 Time-varying analysis of the flow field for 20 degrees PTC cylinder of constant stiffness

60 degrees PTC cylinder as shown in Fig.13. The 60 degrees cylinder displacement and associated power variations exhibit a discernible regularity. During moment **a** as the cylinder progresses from the wave trough to the peak, reaching its maximum displacement, the power attains its zenith. This heightened power coincides with a substantial shedding of vortices on a small scale, with multiple vortices either merging or aligning in series.

Transitioning to moment **b** as the cylinder ascends from the trough to the peak, the power attains an extreme value due to increased velocity. The shedding vortex, influenced by inertia, hovers above the cylinder, while the shed vortices around the cylinder descend in a downward drift.

At moment **c** when the cylinder reaches its peak displacement, intense and relatively symmetric vortex shedding occurs toward the back of the slope. This phenomenon closely resembles the vortex shedding observed behind a stationary cylinder.

Subsequently, at moment **d**, as the cylinder descends to the wave trough, the power diminishes, signifying a braked state. While the vortex shedding is less violent, the flow field beneath the cylinder exhibits heightened velocity in the headward direction.

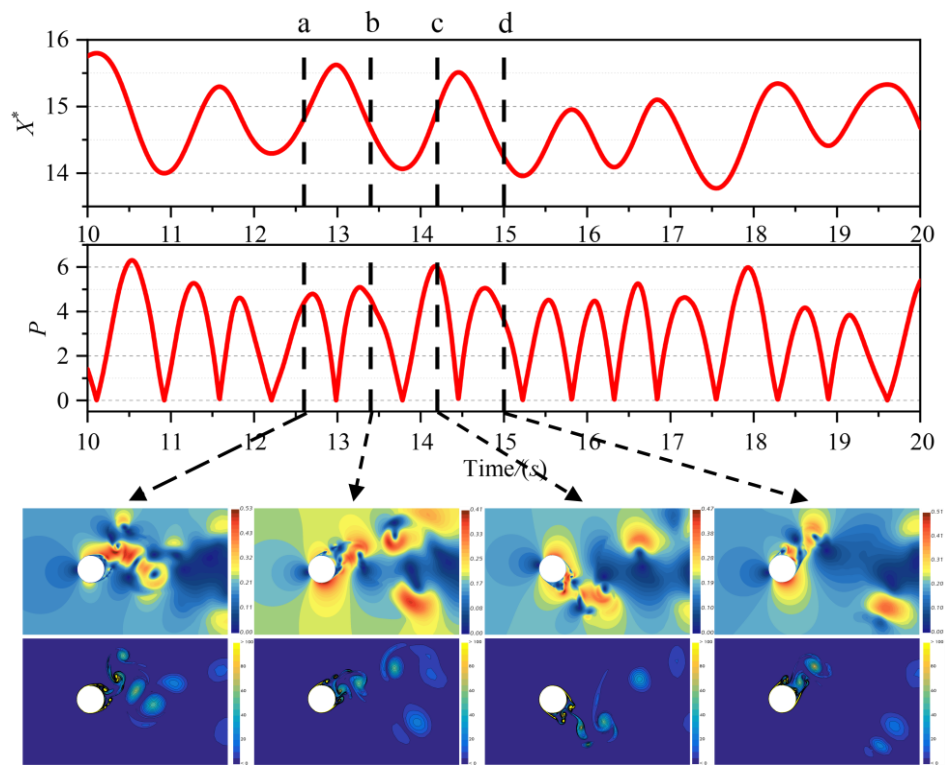

Fig. 13 Time-varying analysis of the flow field for 60 degrees PTC cylinder of constant stiffness

80 degrees PTC cylinder as shown in Fig.14. The 80 degrees displacement of the cylinder, coupled with power fluctuations, exhibits a non-regular pattern influenced by the positioning of the PTC. During moment **a**, as the cylinder resides in the valley, power undergoes a decreasing phase. The vortex configuration behind the cylinder appears chaotic, while the vortices surrounding the recently dislodged or yet-un-dislodged cylinder exhibit a higher degree of symmetry. Intriguingly, a large vortex combines with a smaller one during this phase.

At moment **b**, as the cylinder ascends towards the peak, power levels are relatively high. Vortex shedding, driven by inertia, is characterized by a drifting state. Remarkably, the presence of a large-scale small vortex around the cylinder, coupled with the emergence of two separation points on the cylinder's surface, gives rise to two sandwiched vortices that ultimately coalesce into a single large vortex.

At moment **c** with the cylinder at maximum displacement and power gradually increasing, the rear vortex displays relative symmetry. Two large vortices align symmetrically in the direction of relative motion superposition. Simultaneously, the presence of the PTC induces a collision and merging of vortices at the separation points, forming a turbulent cavity in the wake of the cylinder.

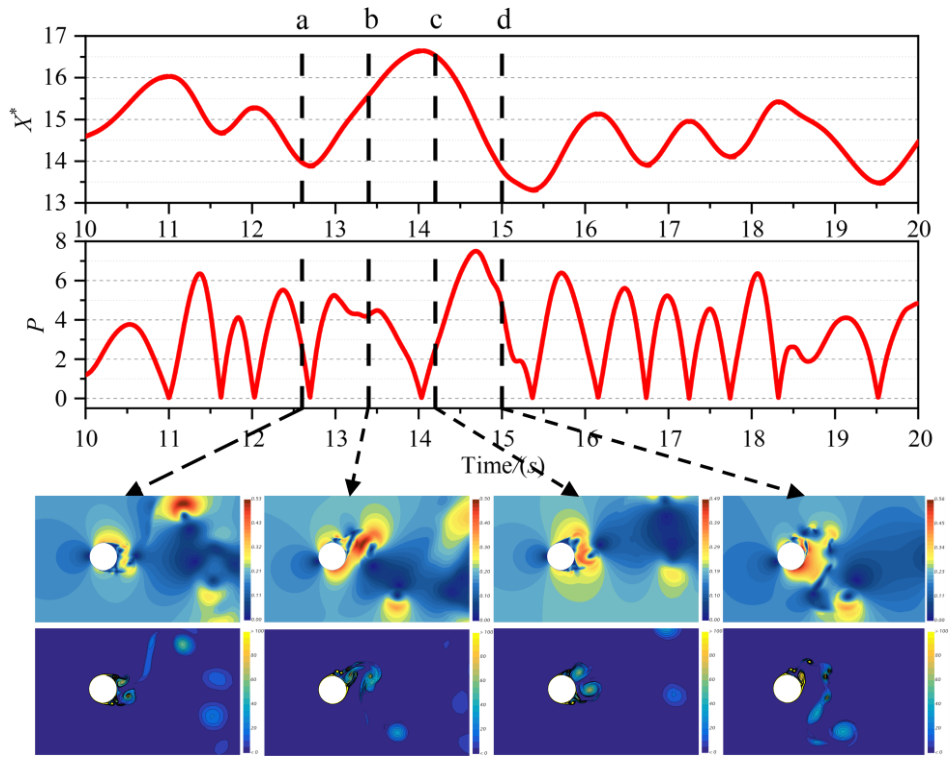

Fig. 14 Time-varying analysis of the flow field for 80 degrees PTC cylinder of constant stiffness

Subsequently, at the moment **d**, when the cylinder displacement reaches a minimal value and power undergoes a significant decrease, the vortex street behind the column reverts to a chaotic state. Small vortices successively engage, but the turbulent cavity observed at the moment **c** undergoes disruption.

In summary, the displacement variations of the bare and 20 degrees PTC cylinders are stable, and the vortex shedding is symmetric and stable. In all cases there is wake drift and vortex asymmetry due to velocity displacement in the y-axis direction. The emergence of the PTC has led to the emergence of large-scale cascading but very small-scale vortices in the wake (tandem small vortices) as well as multilayer flows formed by small vortices in the outer layers and large tail vortices in the classical cylindrical bypass. At the same time, in some cases, there is also a crossover between the first separation point wake and the second separation point wake to form a turbulent cavity with a turbulent flow as shown in Fig.15.

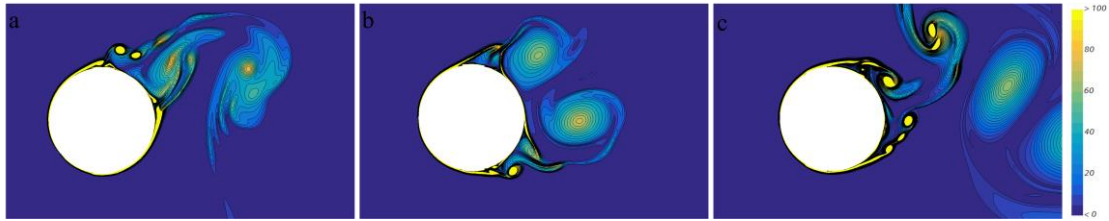

Fig. 15 Extraordinary vortex structure. (a) multilayer flow (b) turbulent cavity (c) tandem small vortices

**3.2 PTC Influence on Periphery**

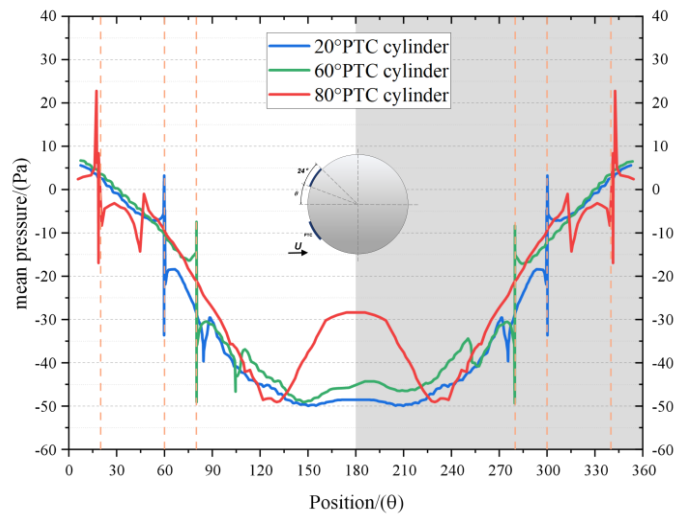

Fig. 16 Expansion of time-averaged pressure distribution on cylindrical surface with different PTC positions

Fig.16 depicts the time-averaged pressure distribution unfolding across the cylinder surfaces at various PTC locations. In configurations featuring PTC, distinct pressure anomalies emerge on the upper surfaces of the cylinders, characterized by a pronounced pressure spike at the leading edge of the PTC and a relatively minor disruption at the trailing edge. While these observations indicate an early onset of localized boundary layer separation, complete boundary layer detachment is not observed.

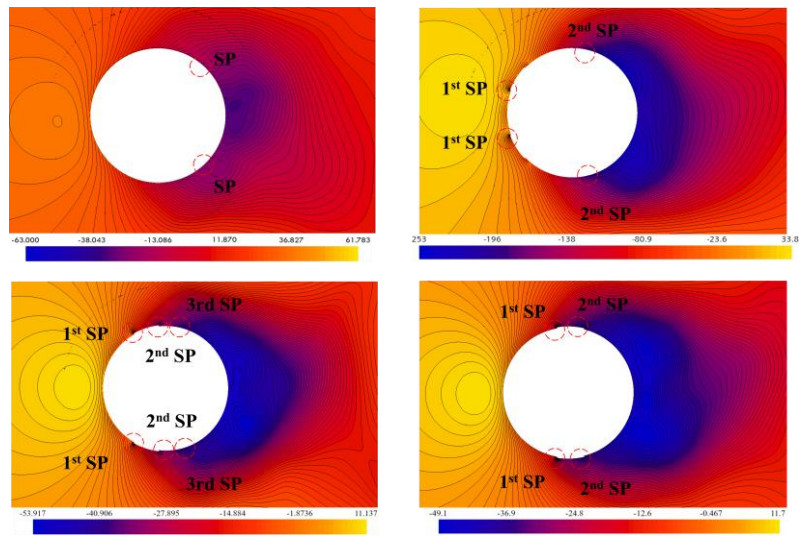

Fig. 17 Pressure gradient cloud around the cylinder with different PTC positions

Because it can be found that the second larger boundary layer appeared in the back half of the cylinder, that is, in the two cases of 20 degrees and 60 degrees PTC, the flow field produced the first and second two boundary layer separation points. In the case of 80 degrees PTC, the PTC is located just over the boundary layer separation point of the normal bare cylinder, and the small vortex generated at the first separation point and the large vortex generated at the second separation point cross and merge, so that there is a sudden reduction of negative pressure and a larger vortex in the wake stream. The situation on the lower surface of the cylinder is symmetrical with the upper surface as a whole. Observations reveal the emergence of a secondary larger boundary layer in the rear half of the cylinder, notably occurring in configurations with PTC positioned at 20 degrees and 60 degrees. In these instances, the flow field induces two distinct boundary layer separation points. Conversely, in the case of an 80 degrees PTC, positioning occurs directly above the boundary layer separation point of the standard bare cylinder. Here, a phenomenon unfolds where the minor vortex generated at the initial separation point intersects and merges with the larger vortex generated at the secondary separation point. This interaction results in a sudden reduction of negative pressure and the formation of a more substantial vortex within the wake stream. Symmetry characterizes the situation on the lower surface of the cylinder in alignment with the upper surface as shown in Fig.17.

### 3.2 PTC and Variable Stiffness Coupled Influence on Harvesting Mechanism

The effects of the coupled PTC and variable stiffness harvesting mechanism on the vibration displacement as well as the real-time power are mainly explored through the analysis of the flow field, taking the SIN variable stiffness form at $w_v/w_n = 0.75$, $U = 0.192$ m/s as an example as shown in Fig.18. They are analyzed from five time points throughout the period, respectively. First of all, from the overall time line, variable stiffness coupled with PTC can greatly increase the displacement and can be maintained in a relatively stable SIN-like periodic motion, the three PTC position has almost no effect on the displacement and frequency of its displacement curve. For the time-calendar power, the power of a part of the time and the bare cylinder are the same, are low-power hibernation section, in the stiffness recovery to the maximum value, the power of the sudden progress to reach the maximum value at the same time. The maximum power is an order of magnitude better with almost no phase difference between the two. However, regarding the relationship between displacement and stiffness, there is a phase difference between the two. When the stiffness returns to half of the constant stiffness, the displacement reaches the maximum. This process is formed by the interactive coupling of spring force, damping, and the force of the fluid on the mechanism. Analyzing the vorticity cloud diagrams at five moments, it can be seen that as the position of the PTC front edge gradually moves backward from the bare cylinder to the 80 degrees PTC, the vorticity shed by the cylinder obviously increases significantly, and the vortex structure is larger and the wake is shed longer. During moment **a** where stiffness is proximal to zero, the power remains dormant, and the displacement is situated in the valley. Notably, vorticity is pronounced, and an extensive wake develops, extending to the lower right due to the combined effects of velocity in the $y$ direction and fluid in the $x$ direction. At moment **b** as stiffness reaches the midpoint, power is poised for a substantial increase, with the displacement reaching its maximum. Observably, vorticity becomes relatively convergent, signifying an imminent turnaround as the spring force surpasses the fluid force induced by vortex shedding. Moving to moment **c** where stiffness and instantaneous power attain their zenith, the displacement is in the diminishing phase. The vortex shedding phenomenon intensifies compared to moment **b** indicating heightened fluid-structure interaction. By moment **d** as stiffness decreases from its maximum, instantaneous power enters a dormant stage, and the displacement reaches the valley. Vortex shedding achieves maximum symmetry relative to the direction of the incoming flow during this moment. Finally, moment **e** mirrors the characteristics of moment **a** marking the transition into the subsequent period of movement.

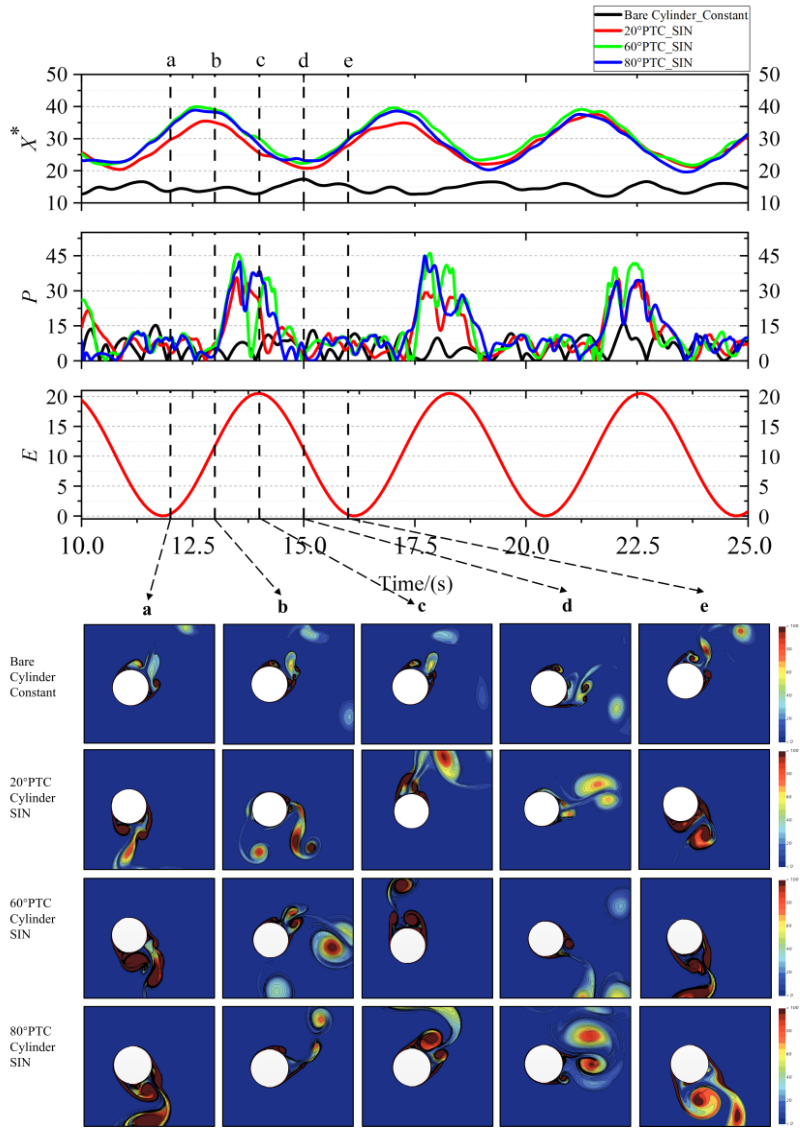

Fig. 18 Time-varying analysis of flow field ($U$=0.192m/s, $w_v/w_n$=0.75)

## 4. Discussion

### 4.1 Energy Harvesting with PTC

Fig.19 shows the motion amplitude and power frequency etc. for different PTC positions at 0.192-0.394m/s (0 represents bare cylinder), and it can be seen that for the amplitude, for all velocities, the 20 degrees PTC is able to increase slightly with respect to the bare cylinder, and is slightly suppressed at both 60 degrees and 80 degrees. As for the power, it can be clearly found that the 20 degrees PTC strongly suppresses its power relative to the bare cylinder, and the power gradually increases from 20 degrees to 60＆80 degrees PTC but is still lower than that of the bare cylinder. This is because although the PTC produces multiple separation points and generates a large number of vortex shedding, but in the process of up and down displacement, vortex drift, small vortex cascade and the merger of large and small vortices, etc., relative to the stable vortex shedding, such an unstable phenomenon, on the contrary, makes the fluid to the energy transfer effect on the cylinder tends to be unstable, inhibiting the increase in its speed of movement, resulting in a decrease in power. There are no large-scale fluctuations in frequency.

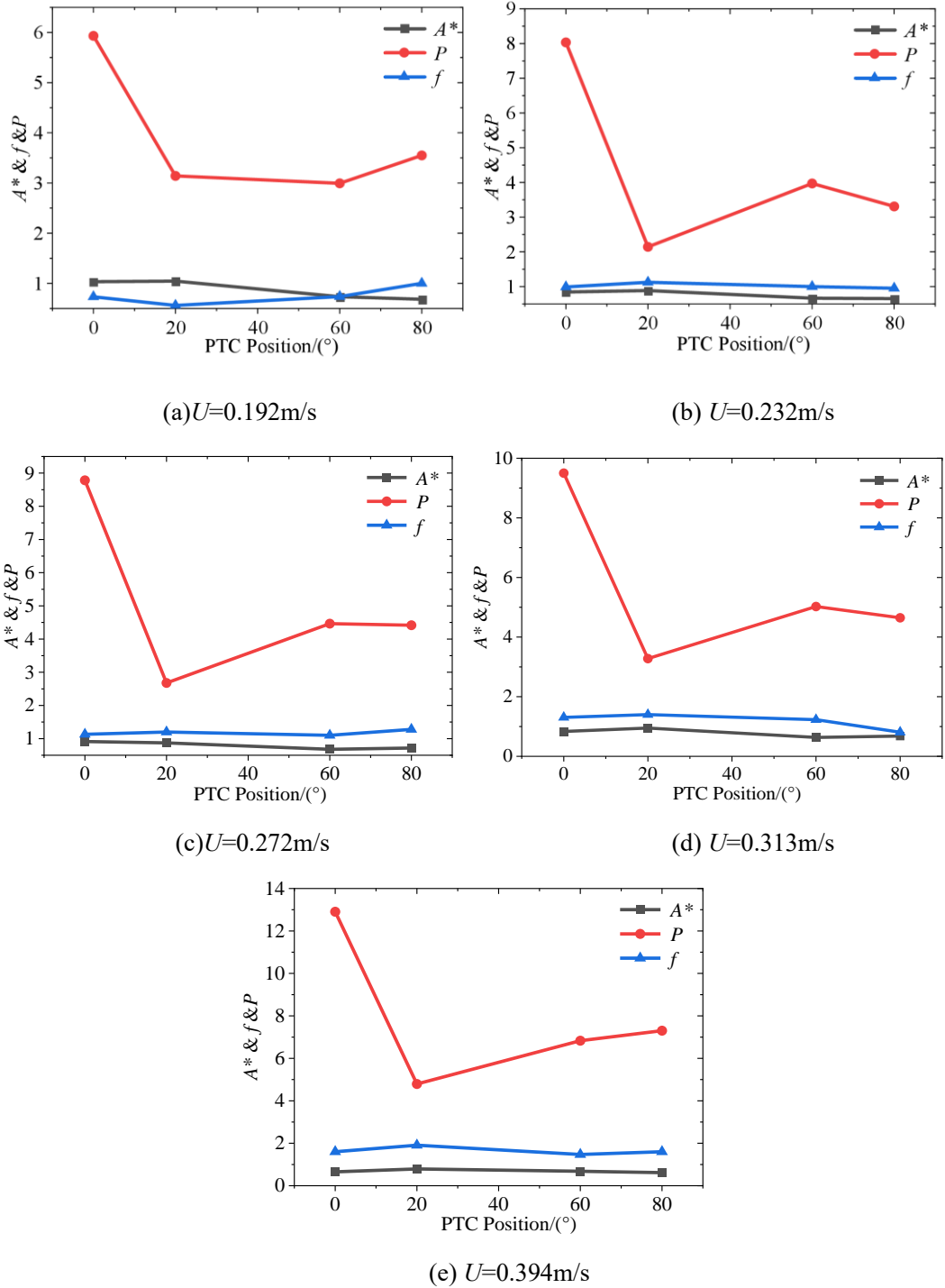

(a) $U$=0.192m/s

(b) $U$=0.232m/s

(c) $U$=0.272m/s

(d) $U$=0.313m/s

(e) $U$=0.394m/s

Fig. 19 Amplitude\power\frequency of motion with different PTC positions at 0.192-0.394 m/s (0 degrees represents no PTC)

The emergence of the PTC clearly breaks the boundary layer flow around the cylinder. A sudden pressure change occurs at all front edge positions where the PTC appears, leading to localized advance boundary layer separation with two or even three separation points. This also has an effect on the motion of the cylinder, the captive power is greatly reduced, and the vibration amplitude increases and then decreases with the backward movement of the PTC position, with a peak at 20 degrees PTC.

**4.2 Energy Harvesting with Variable Stiffness Coupled PTC**

For the various forms of variable stiffness, the sin and trapezoid forms of variable stiffness are chosen which have the best variable stiffness effect. For different frequencies of the SIN form of variable stiffness, the patterns found in Fig.20 (1) SIN form of variable stiffness can effectively increase the cylindrical amplitude (2-11 times), reaching the highest dimensionless amplitude when $w_v/w_n = 0.5$ and improving the captive power up to two times. (2) The overall amplitude decreases throughout the basin as the variable stiffness frequency increases. Frequency increases with increasing variable stiffness frequency. The power does not change much.

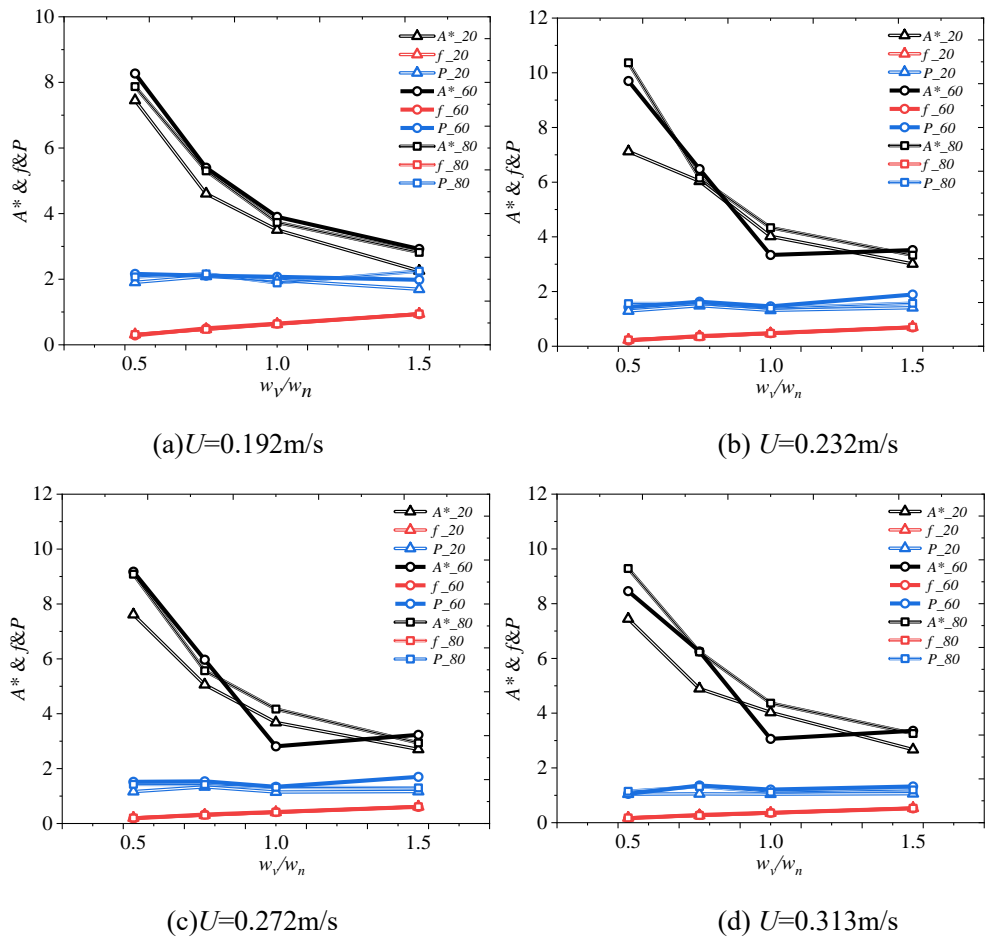

(a) $U=0.192$m/s  (b) $U=0.232$m/s

(c) $U=0.272$m/s  (d) $U=0.313$m/s

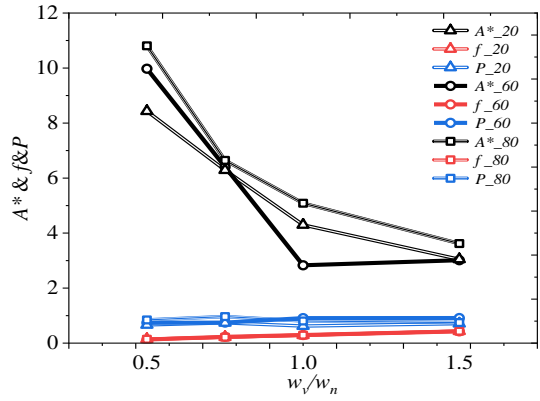

(e) $U$=0.394m/s

Fig. 20 Amplitude/power/frequency of the motion of cylinder with variable stiffness of SIN forms with different frequencies at 0.192-0.394m/s

Fig.21 shows the amplitude/power/frequency of the motion at different frequencies of the trapezoid at 0.192-0.394 m/s, the rules are almost the same except that increasing the amplitude has a worse effect.Fig.22 and Fig.23 show the changes in motion amplitude, power and frequency with speed under different frequency SIN and trapezoid changes of 20 degrees PTC. It can be seen that the overall amplitude of all SIN frequencies does not change much as the flow rate increases, showing up and down fluctuation. However, the energy harvesting power and frequency decrease as the flow rate increases.

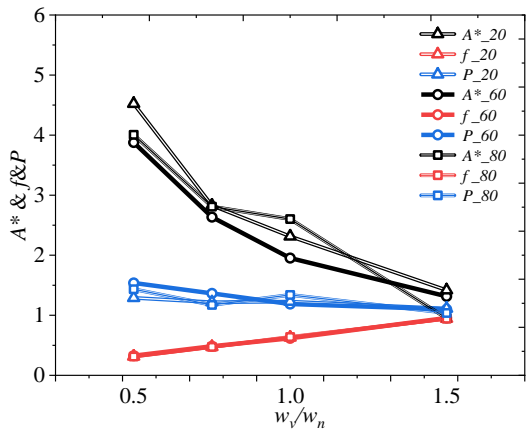

(a)$U$=0.192m/s

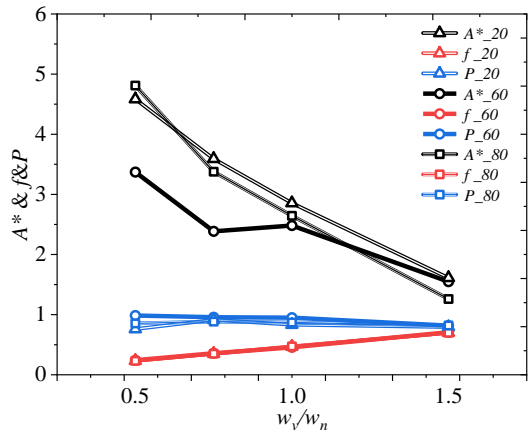

(b) $U$=0.232m/s

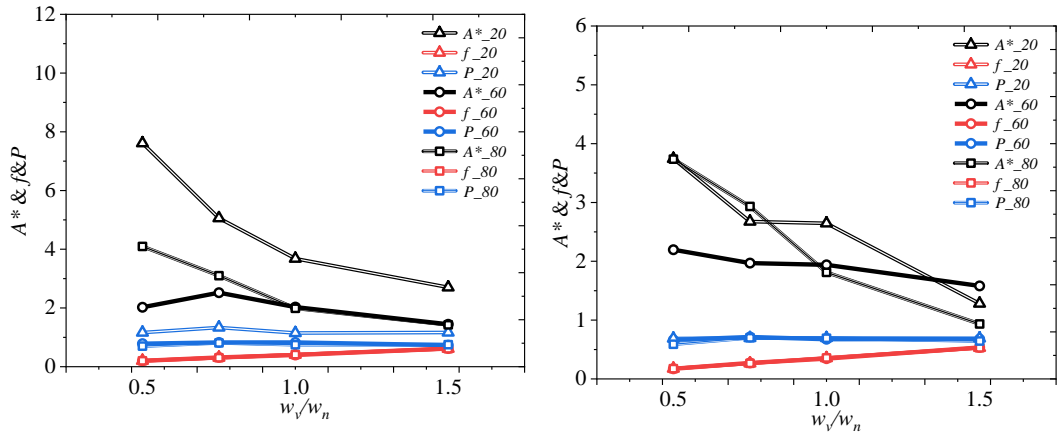

(c) $U$=0.272m/s

(d) $U$=0.313m/s

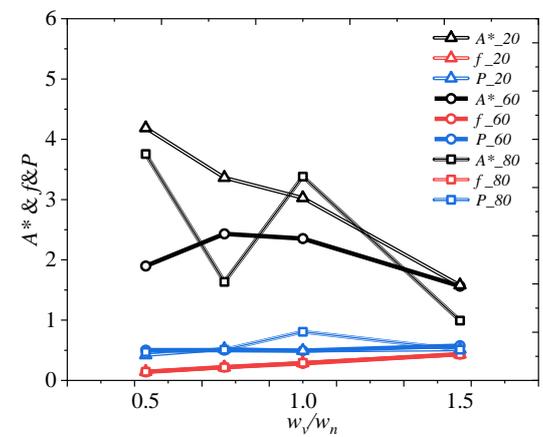

(e) $U$=0.394m/s

Fig. 21 Amplitude/power/frequency of the motion of cylinder with variable stiffness of trapezoid forms with different frequencies at 0.192-0.394m/s

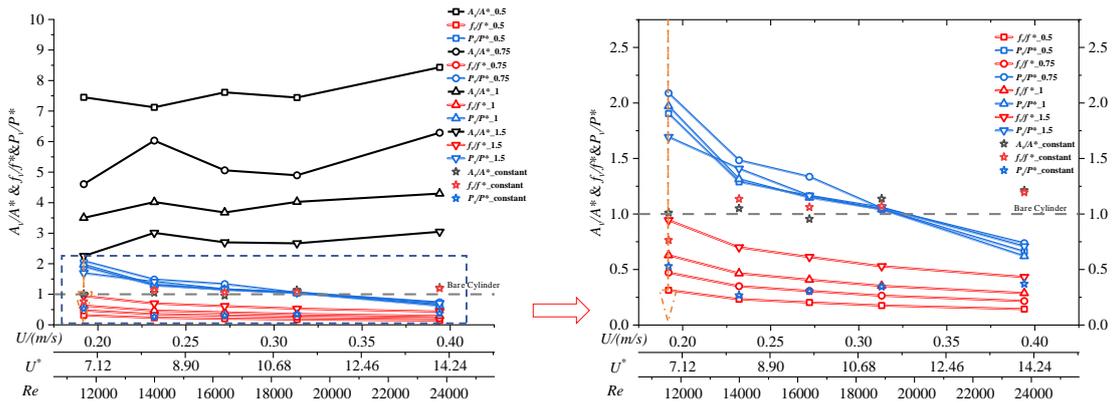

Fig. 22 Amplitude/power/frequency versus velocity for a 20 degrees PTC cylinder stiffened in SIN form at different stiffening frequencies

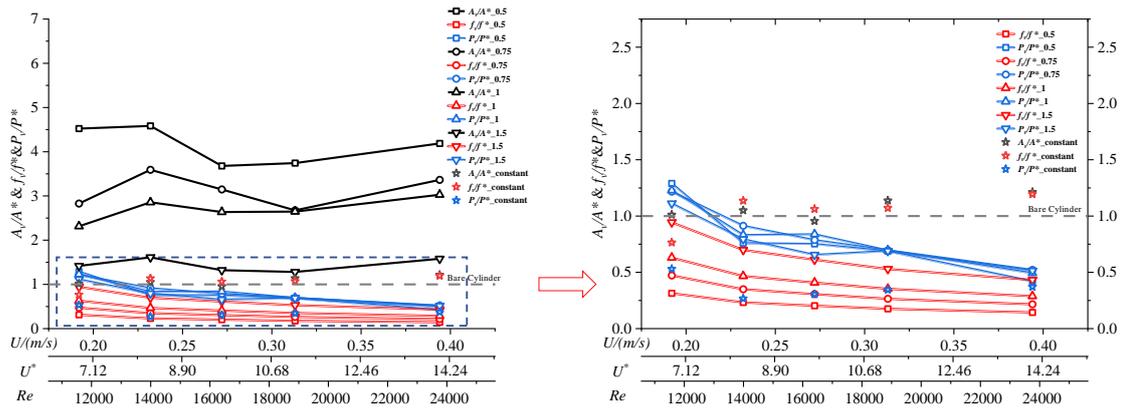

Fig. 23 Amplitude/power/frequency versus velocity for a 20 degrees PTC cylinder stiffened in trapezoid form at different stiffening frequencies

Fig.24 and Fig.25 show the variation of motion amplitude, power and frequency with velocity for 60 degrees PTC with different frequencies of SIN and trapezoid change, it can be seen that the amplitude of all SIN frequencies decreases with the increase of flow velocity, and the gradient is negatively correlated with the frequency of sin change, and the amplitude is almost unaffected by the flow velocity when $w_v/w_n = 1.5$. The power and frequency follow the same pattern. The variable stiffness form of SIN has a better effect on the amplitude and power enhancement of the bare cylinder than the variable stiffness form of the gradient, and the gradient even suppresses the power when exceeding 0.192 m/s. The following figure shows that the amplitude of the 60 degrees PTC decreases with increasing flow velocity overall, and the gradient is negatively correlated with the sin change frequency.

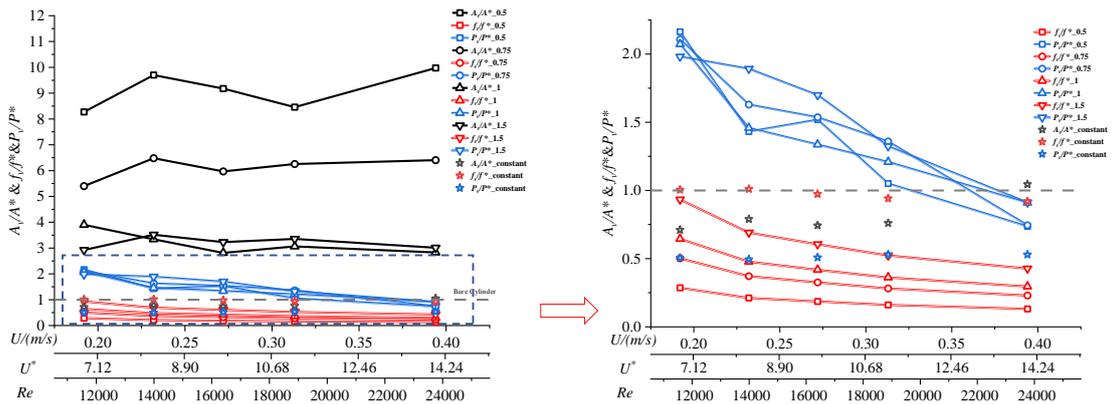

Fig. 24 Amplitude/power/frequency versus velocity for a 60 degrees PTC cylinder stiffened in SIN form at different stiffening frequencies

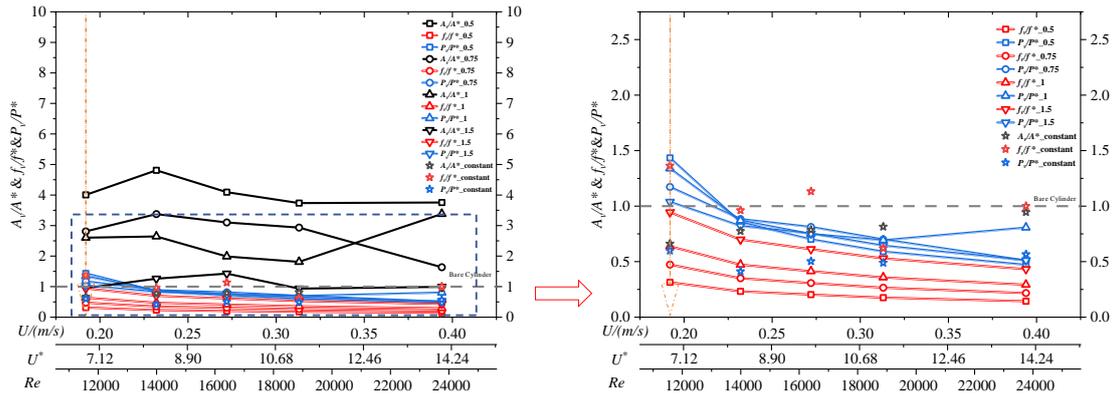

Fig. 25 Amplitude/power/frequency versus velocity for a 60 degrees PTC cylinder stiffened in trapezoid form at different stiffening frequencies

Fig.26 and Fig.27 show the variation of motion amplitude, power and frequency with velocity for 80 degrees PTC with different frequencies of SIN and trapezoid change. It can be seen that the overall amplitude of all SIN frequencies does not change much as the flow rate increases, but shows an overall slightly increasing trend, and the energy harvesting power and frequency decrease as the flow rate increases.

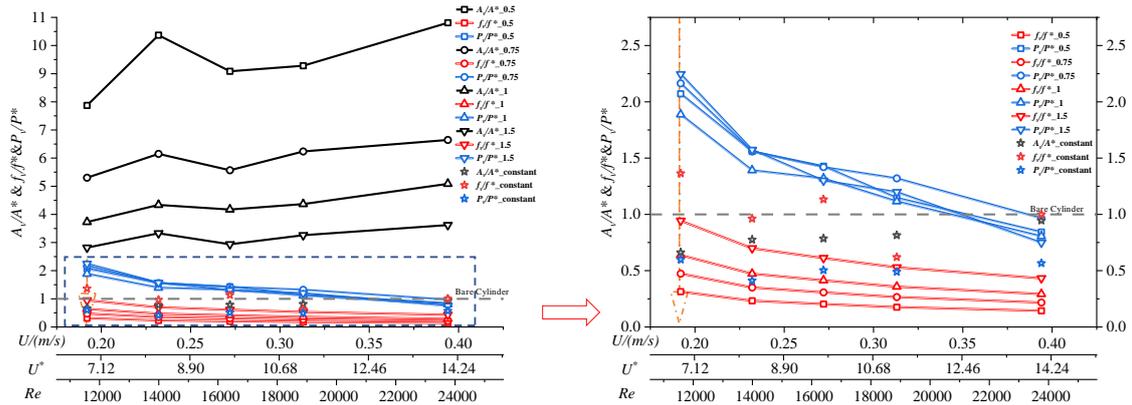

Fig. 26 Amplitude/power/frequency versus velocity for an 80 degrees PTC cylinder stiffened in SIN form at different stiffening frequencies

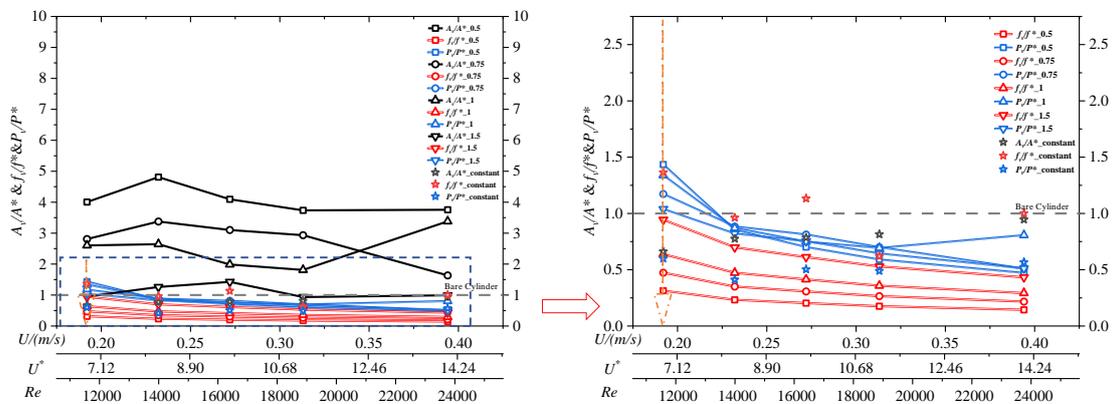

Fig. 27 Amplitude/power/frequency versus velocity for an 80 degrees PTC cylinder stiffened in trapezoid form at different stiffening frequencies

In conclusion, PTC coupled variable stiffness energy harvesting can greatly improve the amplitude of the energy harvesting device and reduce the energy harvesting frequency to realize low-frequency and wide-frequency energy harvesting. With the growth of the flow rate, the flow rate has little effect on the amplitude, the harvesting power and frequency are reduced with the increase of the flow rate, and the lower the flow rate, the better the effect of energy harvesting. In addition, the appearance of PTC does not induce galloping, because the appearance of variable stiffness inhibits this situation.

Fig.28 provides a comprehensive overview of dimensionless displacement time-history curves for both the bare cylinder and the coupled energy harvesting system employing SIN form variable stiffness, across various PTC positions and flow rates investigated. As the flow rate escalates, noticeable fluctuations and shifts are observed in the displacement curves across all cases, reflecting heightened frequency and variability. Nevertheless, a consistent feature persists: the controllable instability induced by the coupling of PTC with SIN variable stiffness.

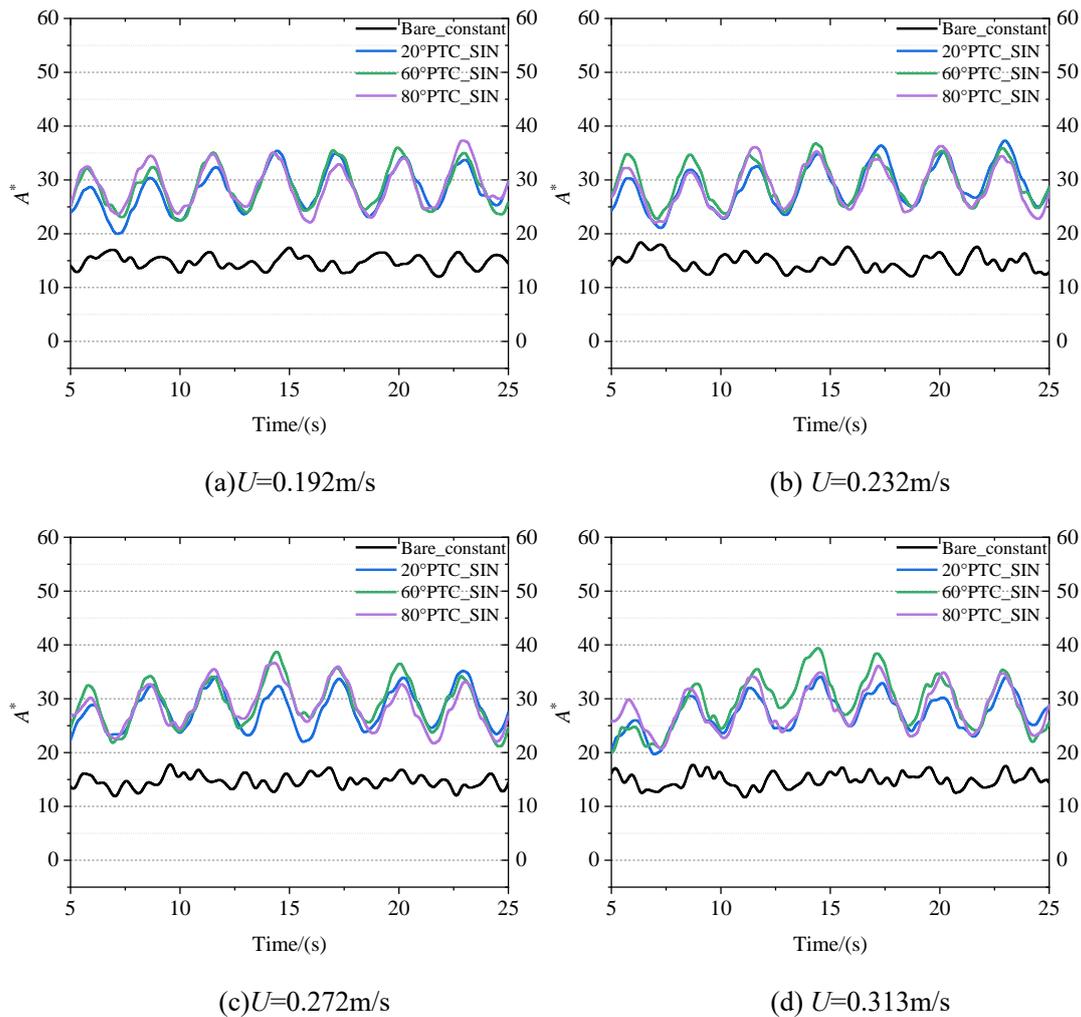

(a) $U$=0.192m/s

(b) $U$=0.232m/s

(c) $U$=0.272m/s

(d) $U$=0.313m/s

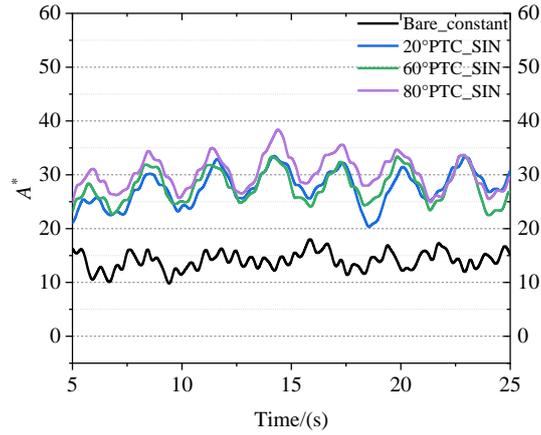

(e) $U$=0.394m/s

Fig. 28 The dimensionless displacement time-alendar curves for the bare cylinder and different PTC positions coupled with SIN variable stiffness for all the flow velocities

Fig.29 shows the spectral analysis diagram of naked cylinder and with passive turbulence control device and PTC-variable stiffness coupling example. It can be found that the PTC device shows a single peak at position 20 degrees, but 60 degrees and 80 degrees show multiple peaks and can achieve energy harvesting for a wider frequency bandwidth. The SIN form and the ladder form variable stiffness dominant frequency finally present slightly lower low-frequency unimodal harvesting energy than the bare cylinder.

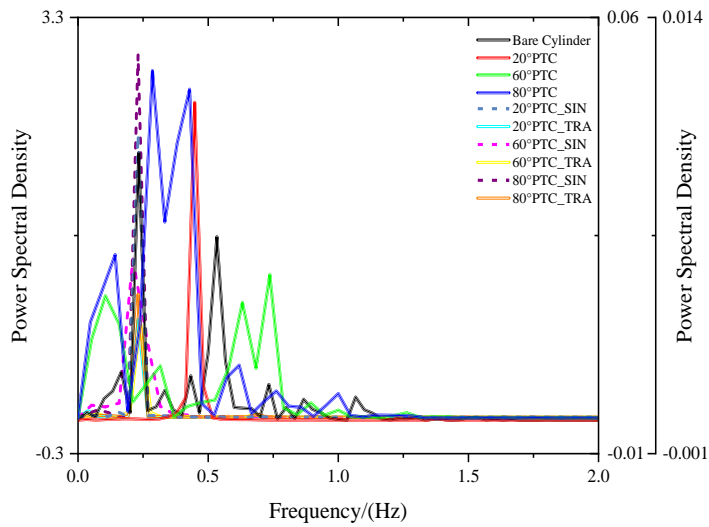

Fig. 29 Spectral analysis diagram ($U$=0.192m/s, $w_v/w_n$=0.75)

In summary, PTC enhance the stability of the system and allow for a broadening of the harvesting frequency, enabling both low frequency and broadband harvesting, and PTC coupled variable stiffness allows the system to be harvested with a combination of high efficiency and stability.

### 4.3 Stabilizing Effect of PTC

Fig 30 illustrates a phase diagram using $U$=0.192m/s and $w_v/w_n$=0.75 as an exemplary scenario. The left panel reveals that the motion trajectory of the bare cylinder tends to diverge, resulting in a larger inner diameter, whereas the cylinder equipped solely with the PTC device tends to converge, yielding a smaller inner diameter. This observation underscores the effectiveness of the PTC device in averting cylinder instability.

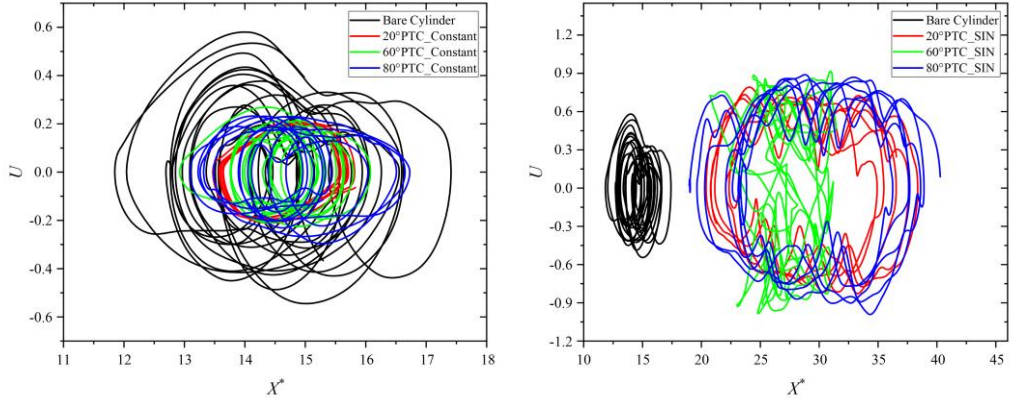

Fig. 30 Comparison of phase diagram ($U$=0.192m/s, $w_v/w_n$=0.75)

Conversely, in the right panel, the coupling effect of PTC and SIN in the form of variable stiffness is evident. Here, the phase diagram displays a larger inner diameter, indicating that the coupling of PTC and SIN in variable stiffness form effectively enhances the vibration amplitude of the energy harvesting device. This coupling mechanism not only mitigates the risk of overall instability, which could jeopardize device integrity and disrupt energy harvesting processes, but also enhances localized instability to augment energy harvesting efficiency.

### 4.4 Energy Harvesting Efficiency

As can be seen from Fig.31, the harvesting efficiency under all the counts shows a decrease with increasing flow velocity, while the gradient gradually decreases, fully exceeding the captive energy efficiency of Xu et al. [21], mostly exceeding the captive energy efficiency of Ding et al [22], and partially exceeding the captive energy efficiency of Sun et al. [23] at 0.192 m/s, as compared with the studies of other researchers. The 80 degrees PTC+SIN variable stiffness form with variable stiffness frequency of 1.5 times the natural frequency has the highest harvested energy efficiency among all the calculations.

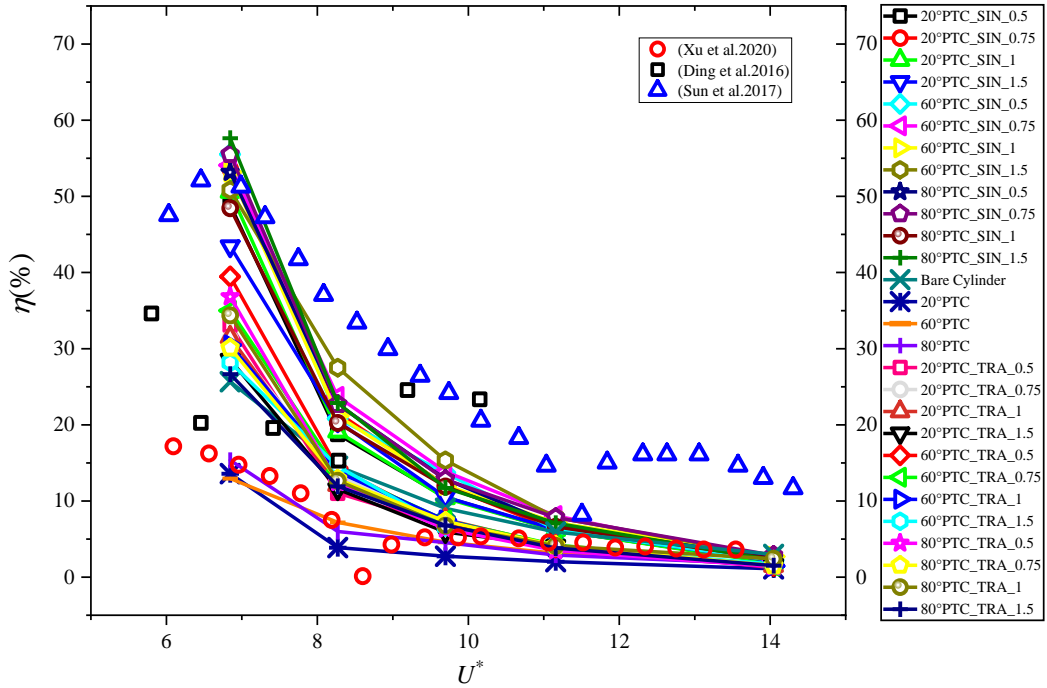

Fig. 31 Energy harvesting efficiency comparison

**4.5 Energy Transfer Mechanism Diagram**

In order to better regulate the two-condition system, the energy transfer mechanism diagram of the process is proposed. The accompanying guiding generalization diagram delineates the interplay of PTC position and two forms of stiffness variations within the FIM, focusing on amplitude and power effects as shown in Fig.32. The classification criteria for amplitude effects include "weak enhancement" for dimensionless amplitude within the range of 1-5 and dimensionless power within 1-2; "strong enhancement" for dimensionless amplitude exceeding 5 and dimensionless power surpassing 2. Conversely, "weak suppression" refers to dimensionless amplitude reduction ranging from 0%-30% and dimensionless power reduction from 0%-50%. Further, "strong suppression " is defined as dimensionless amplitude reduction from 30%-100% and dimensionless power reduction from 50%-100%. In the case of normal stiffness, overall captive energy efficiency remains at a low level. For the PTC position at 20 degrees, there is a weak enhancement in amplitude and a strong suppression in power. At the 60 degrees PTC position, there is a weak suppression in amplitude, and at the 80 degrees PTC position, there is a strong suppression in amplitude and a weak suppression in power. However, both captive energy efficiencies are at a high level. Under SIN variable stiffness, overall harvesting efficiency significantly improves, with all positions exhibiting strong amplitude enhancement and weak power enhancement. Notably, the

80 degrees PTC position yields the most substantial increase in harvesting efficiency. Conversely, the trapezoid form of variable stiffness results in a moderate overall harvesting efficiency. PTC effects on both amplitude and power are characterized by weak enhancement across all positions, with the maximum harvesting efficiency position shifted to the 60 degrees PTC position in advance. This comprehensive analysis serves as a valuable foundation for further research in this domain.

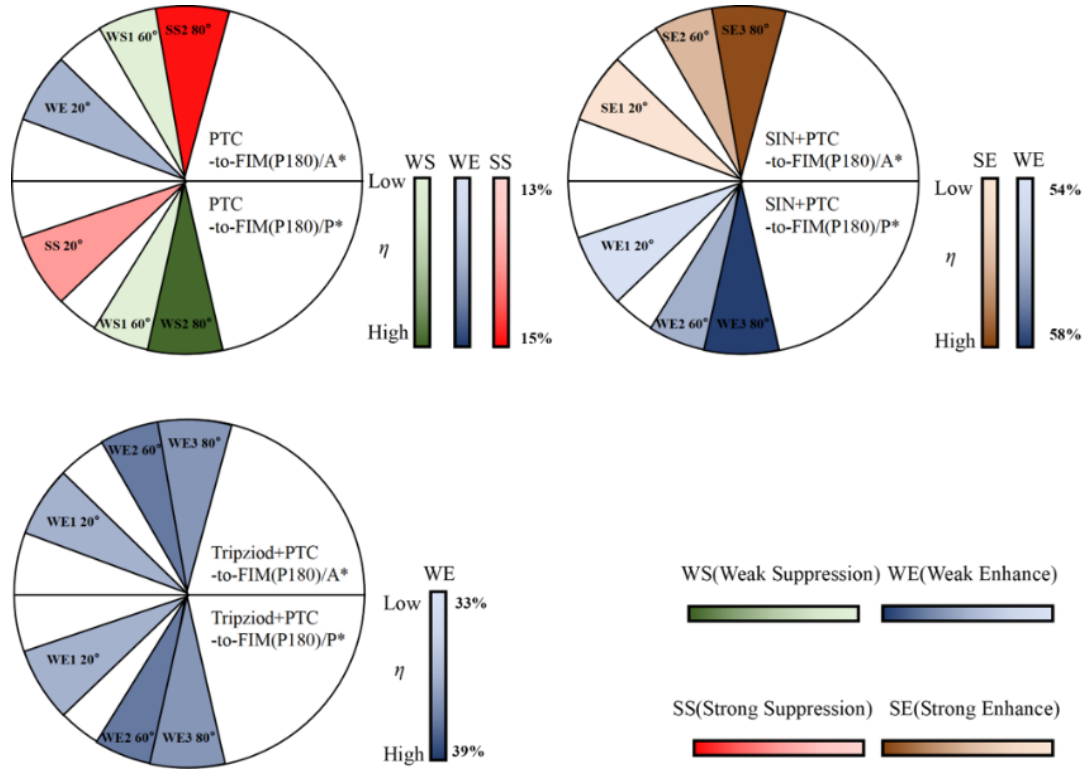

Fig. 32 Mechanism of harvested energy transfer in coupled system

Variable stiffness coupled passive turbulence control achieves low-frequency broadband energy harvesting compared to a bare cylinder. This double-adjusted energy harvesting device not only increases the energy harvesting power, energy harvesting efficiency, and device amplitude, but also makes it possible to harvesting energy in low-speed flows, such as shallow rivers, low-speed sea areas, etc., while increasing its high-efficiency harvesting energy, stability and reliability.

## 5. Conclusion

To address challenges related to narrow bandwidth and inefficient energy harvesting. Our proposition introduces a coupled mechanism integrating boundary layer control and system stiffness modulation inspired by octopus tentacles. Realize broadband and efficient harvesting of energy by interaction between local instabilities and global stability, in which PTC makes global stability by regulating local instability caused by variable stiffness. The following conclusions were obtained:

① The octopus tentacle projection and variable stiffness were transformed into the dual regulation of the passive turbulence control device and the system variable stiffness successfully. An efficient energy harvesting process combining local uncontrolled harvesting device and overall stable energy harvesting is successfully realized. The implementation of PTC promotes global stability through the management of local instabilities due to variable stiffness, resulting in a highly efficient energy harvesting capability.

② Over the range of flow velocities, the amplitude of the cylinder under the coupled system varied little with increasing flow velocity, and the power and frequency, as well as the harvested energy efficiency, decreased with increasing flow velocity. Effective energy harvesting process achieves an effective energy harvesting bandwidth, enhancing the overall energy harvesting efficiency to a maximum of 57%, increasing vibration amplitude by up to 11 times, and raising fluid power output by up to 2 times.

③Propose an energy transfer characteristic map to show the mechanism coupled between boundary layer and stiffness control, offering guiding insights for future related research endeavors. The highest point of harvesting efficiency for the SIN form and step form variable stiffness is at the PTC 80 degrees and PTC 60 degrees positions, respectively, and the SIN form variable stiffness coupled with PTC harvesting is able to increase the harvesting device displacement significantly.

**Acknowledgments**

This work was funded and supported by Chunhui Program (No.: 202201747).

**Reference**

[1] Achenbach, E. 1971. Influence of surface roughness on the cross-flow around a circular cylinder. Journal of fluid mechanics, 46(2), 321-335.
[2] Güven, O., Farell, C., & Patel, V. C. 1980. Surface-roughness effects on the mean flow past circular cylinders. Journal of Fluid Mechanics, 98(4), 673-701.
[3] Nakamura, Y., & Tomonari, Y. 1982. The effects of surface roughness on the flow past circular cylinders at high Reynolds numbers. Journal of Fluid Mechanics, 123, 363-378.
[4] Chang, C. C. J., Kumar, R. A., & Bernitsas, M. M. 2011. VIV and galloping of single circular cylinder with surface roughness at 3.0× 104≤ Re≤ 1.2× 105. Ocean Engineering, 38(16), 1713-1732.
[5] Allen D W, Henning D L. Surface roughness effects on vortex-induced vibration of cylindrical structures at critical and supercritical Reynolds numbers[C]//Offshore Technology Conference. OTC, 2001: OTC-13302-MS.
[6] Blevins, R. D., & Coughran, C. S. 2009. Experimental investigation of vortex-induced vibration in one and two dimensions with variable mass, damping, and Reynolds number.
[7] Vinod, A., Auvil, A., & Banerjee, A. 2018. On passive control of transition to galloping of a circular cylinder undergoing vortex induced vibration using thick strips. Ocean Engineering, 163, 223-231.

[8] Vinod, A., & Banerjee, A. 2014. Surface protrusion based mechanisms of augmenting energy extraction from vibrating cylinders at Reynolds number 3× 103–3× 104. Journal of Renewable and Sustainable Energy, 6(6).

[9] Ding, L., Zhang, L., Kim, E. S., & Bernitsas, M. M. 2015. URANS vs. experiments of flow induced motions of multiple circular cylinders with passive turbulence control. Journal of Fluids and Structures, 54, 612-628.

[10] Ding, L., Zhang, L., Bernitsas, M. M., & Chang, C. C. 2016. Numerical simulation and experimental validation for energy harvesting of single-cylinder VIVACE converter with passive turbulence control. Renewable Energy, 85, 1246-1259.

[11] Park, H., Kumar, R. A., & Bernitsas, M. M. 2013. Enhancement of flow-induced motion of rigid circular cylinder on springs by localized surface roughness at 3× 104≤ Re≤ 1.2× 105. Ocean Engineering, 72, 403-415.

[12] Park, H., Kumar, R. A., & Bernitsas, M. M. 2016. Suppression of vortex-induced vibrations of rigid circular cylinder on springs by localized surface roughness at 3× 104≤ Re≤ 1.2× 105. Ocean Engineering, 111, 218-233.

[13] Park, H., Kim, E. S., & Bernitsas, M. M. 2017. Sensitivity to zone covering of the map of passive turbulence control to flow-induced motions for a circular cylinder at 30,000≤ Re≤ 120,000. Journal of Offshore Mechanics and Arctic Engineering, 139(2), 021802.

[14] Li, N., Park, H., Sun, H., & Bernitsas, M. M. 2022. Hydrokinetic energy conversion using flow induced oscillations of single-cylinder with large passive turbulence control. Applied Energy, 308, 118380.

[15] He, K., Vinod, A., & Banerjee, A. 2022. Enhancement of energy capture by flow induced motion of a circular cylinder using passive turbulence control: Decoupling strip thickness and roughness effects. Renewable Energy, 200, 283-293.

[16] Wu H, Lin Y, Wu Y. Precontrol of short-period motion for a Tension Leg Platform[J]. Journal of Ocean Engineering and Science, 2022.